\documentclass[%
 reprint,
amsmath,amssymb,
aps,
prd,
]{revtex4-2}

\usepackage[utf8]{inputenc}
\usepackage{appendix}
\usepackage{lineno}
\usepackage{duckuments}

\usepackage{placeins}
\usepackage{graphicx}
\usepackage{subfigure}
\usepackage{comment}
\usepackage{chngcntr} 
\usepackage[sort&compress]{natbib}
\usepackage{physics}
\usepackage{xcolor}
\usepackage[normalem]{ulem}
\usepackage{dcolumn}
\usepackage{bm}
\newcommand{\vect}[1]{\boldsymbol{\mathbf{#1}}}
\newcommand{\unit}[1]{\,\mathrm{#1}}
\definecolor{darkmidnightblue}{rgb}{0.0, 0.2, 0.4}

\usepackage[colorlinks=true,linkcolor=darkmidnightblue,urlcolor=darkmidnightblue,citecolor=black]{hyperref}
\usepackage[all]{hypcap} 

\begin{document}

\title{Energy estimation of direct and reflected cosmic-ray events aboard balloon-borne radio detectors}

\author{Sergio Cabana-Freire}
\email{sergio.cabana.freire@usc.es}
\author{Jaime \'Alvarez-Mu\~niz}%
\email{jaime.alvarez@usc.es}
\affiliation{
 Instituto Galego de Física de Altas Enerxías (IGFAE) \& Universidade de Santiago de Compostela \\
 15782 Santiago de Compostela, Spain.
}%

\author{Matias Tueros}
\email{tueros@fisica.unlp.edu.ar}
\affiliation{
 Instituto de Física La Plata, CONICET-UNLP \\
Diagonal 113 entre 63 y 64, La Plata, Argentina.
}%

\begin{abstract}

Balloon-borne radio detectors offer a highly effective approach to monitoring immense volumes of the Earth's atmosphere and ice for ultra-high-energy cosmic rays and neutrinos. Because these payloads measure radio emission from a single, highly elevated vantage point, reconstructing the air-shower energy using traditional methods is not possible. In this work, we present a comprehensive, simulation-based energy reconstruction framework tailored to the unique cosmic-ray event geometries observed by balloon payloads. For the first time, we adapt and apply this methodology to \textit{direct} atmosphere-skimming cosmic-ray showers, accounting for their highly asymmetric development in rarefied air. Furthermore, we provide a major update to the energy reconstruction of \textit{reflected} downward-going showers by incorporating spherical reflection and ray de-focusing mechanics directly into simulations. We rigorously validate the method across different event geometries, demonstrating a baseline energy resolution of $8-11\%$. Finally, we quantify the systematic effects of finite detector pointing resolution, unknown primary mass composition, and signal-to-noise ratio, establishing a robust framework for the analysis of cosmic-ray events in next-generation balloon missions such as PUEO and POEMMA-Balloon with Radio.

\end{abstract}

\maketitle


\section{Introduction}\label{sec:Intro}

The detection of ultra-high-energy (UHE, $>10^{17}\unit{eV}$) cosmic rays, as well as searches for UHE neutrinos of astrophysical origin, relies on  monitoring very large volumes of target material where they can interact and produce showers of secondary particles. In particular, the detection of the radio pulses produced by in-air or in-ice particle showers has become a ubiquitous approach to the detection of ultra-high-energy neutrinos, that enables a cost-effective deployment of instrumentation monitoring large interaction volumes 
\cite{Allison:2011wk, GRAND:2018iaj, RNO-G:2020rmc, Horandel:2025Km, Abbasi:2025PO, BEACON:2025qcq, HERON}.

Many of the experiments using the radio technique consist of arrays of antennas deployed over large areas and operating over extended periods of time, to accumulate the exposure needed to detect or constrain the flux of UHE neutrinos expected in models. Balloon-borne radio detectors such as ANITA \cite{ANITA:2008mzi} and PUEO \cite{PUEOWhitePaper} follow an alternative approach to obtain sensitivity to UHE neutrinos interacting near the Earth's surface. By placing a densely packed array of antennas at altitudes of $\sim35-40\unit{km}$ in a month-long flight, the long operational lifetime and physical size of large antenna arrays are traded for a much larger instantaneous field of view. Other balloon- and satellite-borne initiatives sharing this approach are POEMMA \cite{POEMMA:2020ykm} and its preparatory flights, EUSO-SPB2 \cite{AdamsJr:2026qlh} and POEMMA-Balloon with Radio (PBR) \cite{POEMMA_BR, PBRTeam:2026wxs}.

The reduced size of balloon-borne payloads restricts the available information on the radio emission produced by in-air or in-ice particle showers. While ground-based radio arrays are able to measure signals at different positions around shower axis to reconstruct the event direction and the emitted radiation pattern \cite{PierreAuger:2016vya, Mitra:2020mza, Schluter:2022mhq, Guelfand:2025goo, Zhang:2025rzp}, balloon-borne detectors typically measure the radio emission at a single location around the shower axis, with several antennas in proximity to each other. This allows to determine the incoming direction of the radio pulse with sub-degree accuracy \cite{Romero-Wolf:2014pua}, but a complete interferometric reconstruction of the shower geometry, as that performed in \cite{Schoorlemmer:2020low, PierreAuger:2025jaw},  is more challenging due to the reduced baseline between antennas. Under these limitations, the direction of the particle cascade cannot be unambiguously reconstructed. As a result, the polarization and frequency content of the measured radio signal as well as its incoming direction, become key to obtain information about the origin and energy of the detected events.

Searches for radio signals with the incoming directions and polarizations expected for neutrinos interacting in Antarctic ice did not reveal any candidate across the four flights of the ANITA experiment \cite{ANITA:2008wdk, ANITA:2010hzc, ANITA:2018vwl, ANITA:2019wyx}. On the other hand, multiple cosmic-ray events were recorded, exhibiting the impulsivity, frequency spectra and polarization expected for air showers emitting in radio-frequencies through the geomagnetic mechanism \cite{ANITA:2010ect, ANITAEnergyFlux, ANITAIII, ANITAIV}. In the context of the ANITA experiment, these events were subdivided into two classes: \textit{direct} events, characterized by incoming directions pointing above the ground, and produced by cosmic-ray air showers skimming the Earth's atmosphere; and \textit{reflected} events, with incoming directions pointing to the ground and exhibiting a characteristic inversion of the signal polarity with respect to that of direct events. These two features are interpreted as signatures of radio emission produced by downward-going cosmic ray air showers, undergoing a reflection on the ice surface of the Antarctica before reaching the payload. Cherenkov emission produced by direct cosmic-ray events has also been observed during the EUSO-SPB2 balloon flight \cite{AdamsJr:2026qlh}.

The diffuse cosmic-ray flux represents an almost unavoidable background in radio experiments focused on UHE neutrino detection. Rather than being merely a background, cosmic-ray signals provide an invaluable sample for developing and validating event reconstruction algorithms. In addition, UHE cosmic-ray events can yield important information on the energy threshold for detection, instrument response and signal properties. 

Due to the reduced detection threshold in comparison to ANITA, experiments like PUEO and PBR are expected to record a substantially larger number of cosmic-ray events during their flights. In this context, this paper provides an advancement in the reconstruction of cosmic-ray energies from ballon-borne radio observations. We present an updated energy reconstruction method, based on that introduced in \cite{ANITAEnergyFlux}, together with its first rigorous validation using dedicated simulations. Moreover, we extend the energy reconstruction to direct events, whose energies were not estimated in \cite{ANITAEnergyFlux} because dedicated simulations, and hence an accurate modeling of their radio signal properties, were not yet available.

This paper is structured as follows: Section \ref{sec:method} contains an overview of the energy estimation method, that can be applied both to direct and reflected cosmic-ray events recorded aboard balloon-borne radio detectors. The application of the method to direct cosmic-ray events is presented in Section \ref{sec:direct}, where a systematic evaluation of its performance is carried out. In Section \ref{sec:reflex}, the energy reconstruction method is applied to reflected cosmic ray events with a similar approach to \cite{ANITAEnergyFlux}, now including significant updates at the simulation level and a first evaluation of the performance against simulated events. Finally, in Section \ref{sec:conclusions} we conclude the paper.


\section{Energy reconstruction method}\label{sec:method}

The cosmic-ray energy reconstruction method presented in this work follows a similar approach to that employed in \cite{ANITAEnergyFlux}. The method, reviewed in the following, exploits the dependence of the frequency spectrum of the radio signal on the position of the observer around shower axis, together with the knowledge of the incoming direction of the pulse and the position of the payload.

\begin{figure}
    \centering
    \includegraphics[width=\linewidth]{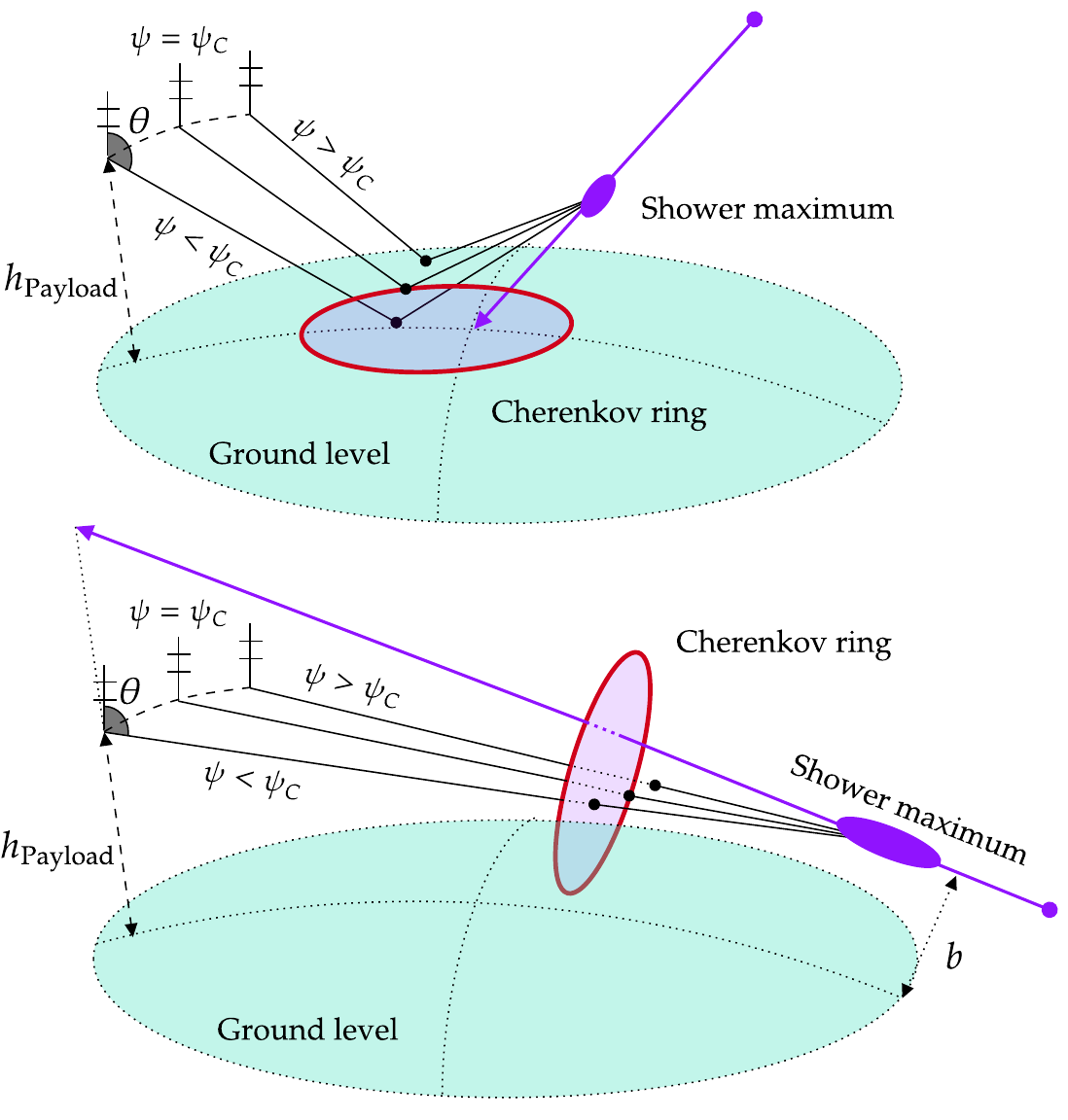}
    \caption{Sketch of the shower geometry in \textit{reflected} (top) and \textit{direct} (bottom) cosmic-ray events detected by balloon-borne radio detectors. The shower axis is represented with a purple arrow with the shower maximum assumed as the effective source of most of the radio emission. Top: The radio emission from a downward-going air shower (black lines) reflects off the Earth's surface before reaching the detector. A \textit{slice} of the Cherenkov cone on the ground is shown, with the Cherenkov ring (see text for definition) in red. The incoming direction of the radiation ($\theta$ in the sketch) points back to the ground. Bottom: Radio emission from an atmosphere-skimming air shower whose axis does not intersect ground, characterized by an impact parameter $b$. A \textit{slice} of the Cherenkov cone in a plane perpendicular to the shower axis is represented in red. The incoming direction of the radiation points above the horizon in these events.
    }
    \label{fig:geometry}
\end{figure}

\begin{figure*}
    \centering
    \includegraphics[width=\linewidth]{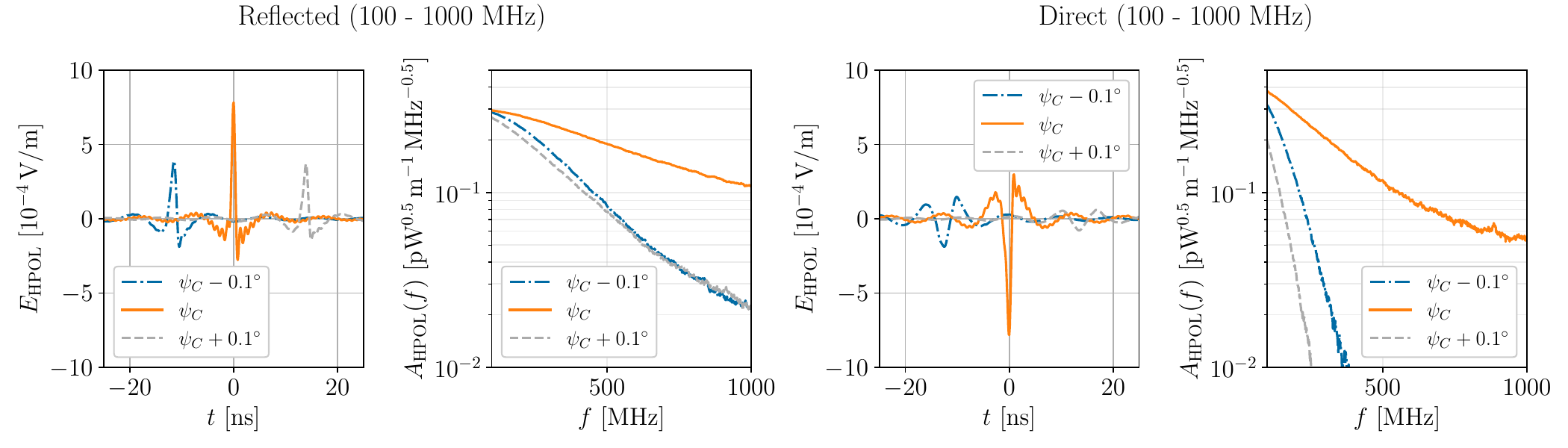}
    \caption{Simulated radio-pulses along the horizontal polarization, i.e. parallel to ground at the position of the detector, observed by detectors placed at $36\unit{km}$ above sea level, around the angular position $\psi_C$  where the high frequency content is maximal. Left panels: Radio pulses in time and frequency domain induced by a downward-going proton shower after reflecting on a perfectly-smooth spherical surface of ice, with an incidence angle $\alpha=74.86^\circ$. Right panels: Radio pulses produced by an atmosphere-skimming proton cascade. The arrival direction of radio pulses $\theta$ points above the horizon. A rectangular filter in the $100-1000\,\mathrm{MHz}$ range has been applied to all time traces. Amplitudes and arrival times have been arbitrarily scaled. The sketches in Fig.\,\ref{fig:geometry} are representative of the shower geometries adopted in the simulation of these two events performed with the \textsc{ZHAireS} \cite{Reflex} and \textsc{ZHAireS-RASPASS} Monte Carlo codes \cite{RASPASS_Radio}.}
    \label{fig:pulses}
\end{figure*}

The macroscopic extent of air showers, typically ranging from hundreds of meters to several kilometers, limits the maximum frequency at which the radio emission from the entire cascade can superpose coherently. As the observation frequency increases beyond $\order{100\,\mathrm{MHz}}$, the coherence of the radio signal diminishes. Simulations of radio signals produced by cosmic-ray air showers \cite{RadioAirShowers} reveal that the frequency spectrum of the signal amplitude $A(f)$ can be represented, in this frequency range, with an exponential function,
\begin{equation}
    A\left(f\right) = A_0 \; e^{\gamma\left(f-f_0\right)}\,,
    \label{eq:expmodel}
\end{equation}
\noindent where the exponent $\gamma < 0$ and throughout this work we set $f_0 = 300\unit{MHz}$. This behavior has been observed both in reflected and direct radio signals recorded by ANITA \cite{ANITA:2010ect, ANITAEnergyFlux}. The geometries of such events are sketched in Fig.\,\ref{fig:geometry}, while corresponding simulated examples of radio pulses can be seen in Fig.\,\ref{fig:pulses}.

The frequency content of the received radio signal depends on the position of the observer around shower axis, as can be seen in Fig.\,\ref{fig:pulses}. At a certain angle $\psi_C$ with respect to the shower direction, the emission from the whole longitudinal development reaches the observer almost simultaneously, increasing the high-frequency content or \textit{coherence} of the radio signal \cite{RadioAirShowers}. Observers at this angular position $\psi_C$, register the largest, i.e. less negative, values for the spectral slope $\gamma$ in Eq.\,\eqref{eq:expmodel}. This region defines the \textit{Cherenkov ring} around shower axis where the radio signal has maximum amplitude (as sketched in Fig.\,\ref{fig:geometry}). As the observer moves away from the Cherenkov ring, the increased time delays between the emission from different regions of the shower reduce the high-frequency content of the signal.

\begin{figure}
    \centering
    \includegraphics[width=\linewidth]{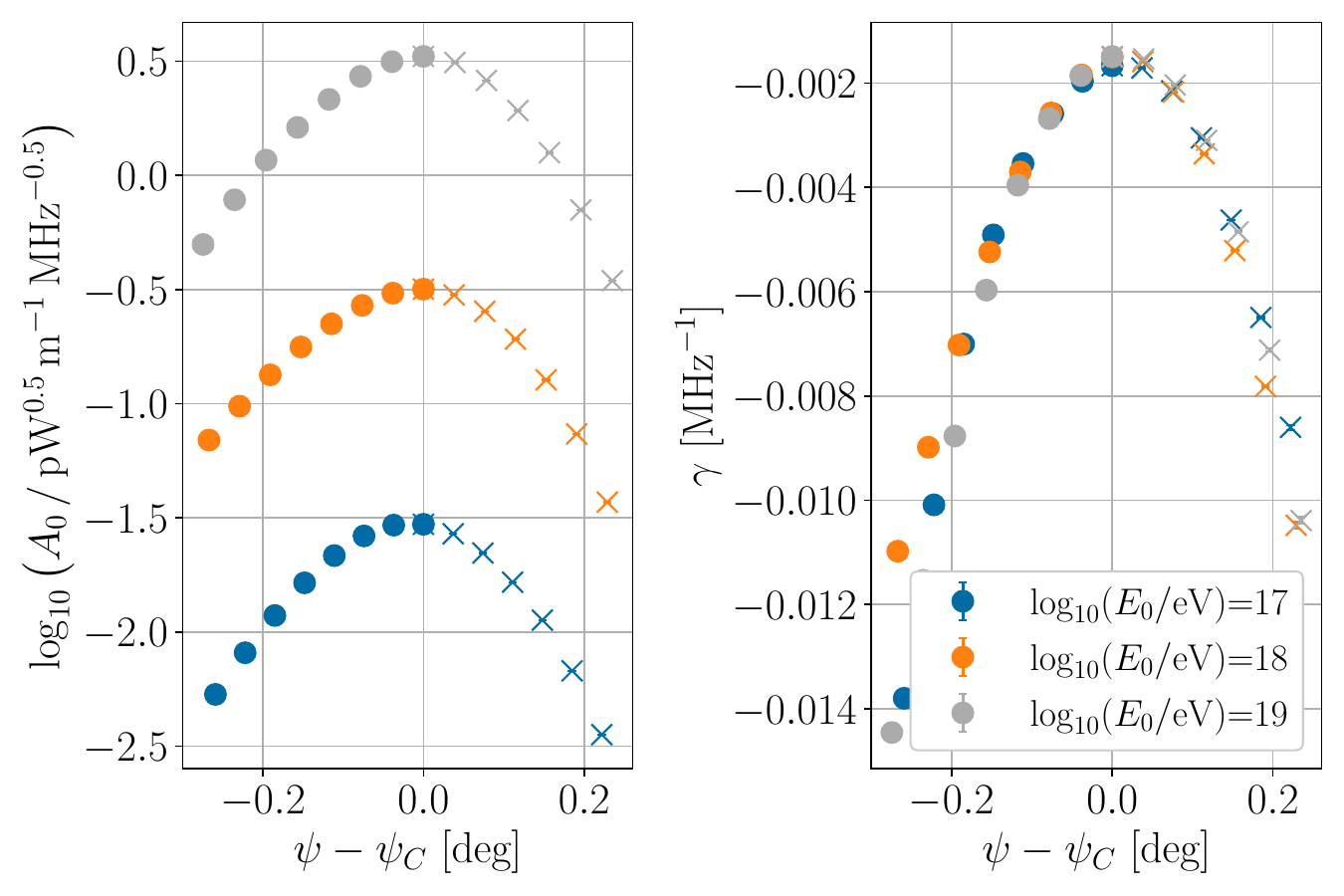}
    \caption{Left panel: Values of the spectral amplitude parameter $A_0$ in Eq.\,\eqref{eq:expmodel} as a function of the angular position of the observer $\psi$ (shown with respect to the angle $\psi_C$ where the high frequency content of the signal is maximal), as obtained in simulations of downward-going proton showers for three primary energies. Right panel: Same as left panel for the spectral slope parameter $\gamma$. The shower geometry and position of the observers are sketched in the top panel of Fig.\,\ref{fig:geometry}. The incidence angle of the radio signal on the ground is fixed to $\alpha=74.86^\circ$.  Spectral fits were performed in the $300-1000\,\mathrm{MHz}$ range. Observers \textit{inside} ($\psi<\psi_C$) or \textit{outside} ($\psi>\psi_C$) the Cherenkov ring are represented with different markers.}
    \label{fig:curves}
\end{figure}

The dependence of the spectral \textit{amplitude} $A_0$ and \textit{slope} $\gamma$ as a function of the primary energy and observer angular position \footnote{ Measured with respect to the shower maximum $X_{\rm max}$} is shown in Fig.\,\ref{fig:curves}, for the case of the reflected emission produced by downward-going proton showers.

The spectral amplitude parameter $A_0$ in Eq.\,\eqref{eq:expmodel} depends both on the primary energy, which determines the number of emitting particles, and on the off-axis observation angle $\psi$. As shown in the left panel of Fig.\,\ref{fig:curves}, $A_0$ decreases as the observer moves away from the Cherenkov angle $\psi_C$, where signal coherence is maximal. Consequently, the signal amplitude alone cannot yield the primary energy without knowledge of $\psi$.

Balloon-borne payloads measure only the arrival direction of the radio signal \cite{Romero-Wolf:2014pua, ANITAIII}, which is insufficient to uniquely determine the shower axis geometry and thus the angle $\psi$ at which the signal is received. As illustrated in Fig.\,\ref{fig:showerselecdirect}, because the radio emission is beamed within a narrow cone, the same arrival direction $\theta$ could be consistent with observations at various off-axis angles.

This ambiguity can be broken using the spectral slope $\gamma$, which characterizes the macroscopic coherence of the signal. As demonstrated with simulations of both downward-going and atmosphere-skimming events using \textsc{ZHAireS} \cite{Reflex} and \textsc{ZHAireS-RASPASS} \cite{RASPASS_Radio}, the value of $\gamma$ is governed almost entirely by $\psi$, with only a marginal dependence on primary energy (Fig.\,\ref{fig:curves}, right panel). This relationship makes it possible to express $A_0$ as function of $\gamma$ rather than $\psi$ which is unknown.

\begin{figure}
    \centering
    \includegraphics[width=\linewidth]{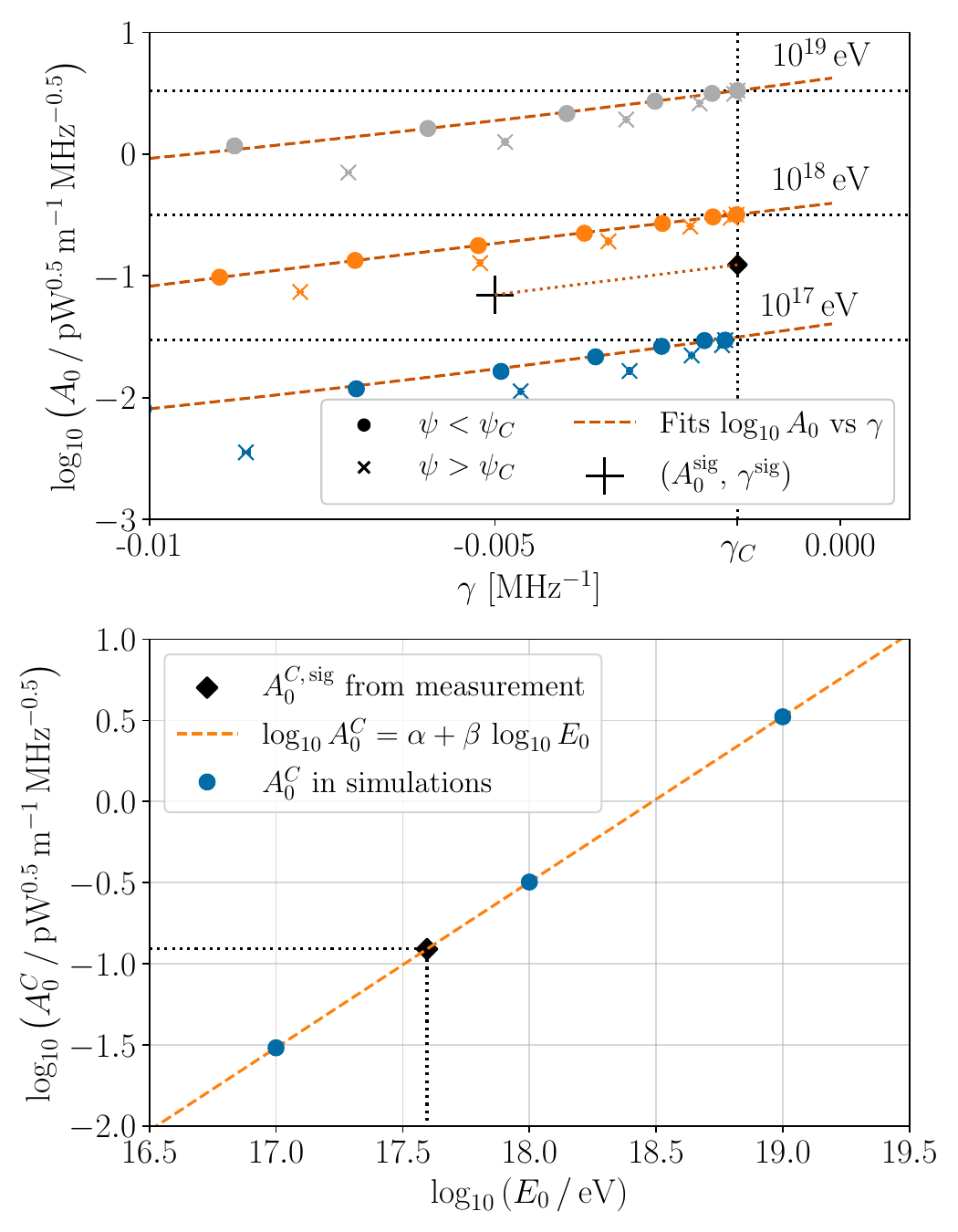}
    \caption{Top panel: Calibration plot showing the spectral amplitude $A_0$ in Eq.\,\eqref{eq:expmodel} as a function of the spectral slope $\gamma$ for the same geometry as in Fig.\,\ref{fig:curves}. A vertical dotted line indicates the average spectral slope $\gamma_C$ at the position where the signal coherence is maximal. Horizontal dotted lines indicate the spectral amplitude at the Cherenkov angle $A_0^C$ for three different primary energies. Fits to Eq.\,\eqref{eq:cal_line} are shown with dashed lines, for observers inside the Cherenkov cone. A mock measurement $(A_0^{\rm sig}, \,\gamma^{\rm sig})$ is shown with a cross symbol, including the extrapolation of the amplitude to the Cherenkov angle (orange dotted line ending at the black square). Bottom panel: Linear fit to the spectral amplitude at the Cherenkov angle $A_0^C$ as a function of primary energy in simulations. The extrapolated amplitude at the Cherenkov angle $A_0^{C,\,{\rm sig}}$ for the mock event (black square) allows to estimate its primary energy. Error bars are smaller than markers in all plots. Only three primary energies (blue circles) are shown for clarity.}
    \label{fig:calibration}
\end{figure}

Combining the information in the left and right panels of Fig.\,\ref{fig:curves}, the amplitude parameter $A_0$ is shown as a function of $\gamma$ in the top panel of Fig.\,\ref{fig:calibration}.
For fixed primary cosmic-ray energy $E_0$, the amplitude parameter $A_0$ grows with $\gamma$, as the observer gets closer to the angle $\psi_C$ where coherence is maximal. Since the lateral distribution of the radio emission is in general not symmetrical around $\psi_C$, the dependence $A_0(\gamma)$ can be slightly different for observers \textit{inside} ($\psi<\psi_C$) or \textit{outside} ($\psi>\psi_C$) the Cherenkov ring. This effect is visible in the top panel of Fig.\,\ref{fig:calibration}.

Yet, the growth of the amplitude $A_0$ with $\gamma$ for observers on either side of the Cherenkov ring, is similar for all primary energies. This dependence can be modeled as,
\begin{equation}
    \log_{10}A_0 = \log_{10}A_0^C + b\left(\gamma - \gamma_C\right) + c\left(\gamma - \gamma_C\right)^2\,,
    \label{eq:cal_line}
\end{equation}
\noindent where $\gamma_C = \gamma\left(\psi_C\right)$ is the maximal spectral slope found in simulations, and $A_0^C=A_0(\psi_C)$ represents the spectral amplitude of the maximally coherent signal at $\psi=\psi_C$. The parameters $b$, $c$ and $\gamma_C$ in Eq.\,\eqref{eq:cal_line} show no significant dependence on primary energy, which affects only the overall normalization of the radio amplitudes represented by the factor $\log_{10}A_0^C$ in Eq. \eqref{eq:cal_line}. Therefore, the spectral amplitude $A_0^C$ at the angle $\psi_C$ scales linearly with the primary energy. This is shown in the bottom panel of Fig.\,\ref{fig:calibration}. A simple scaling relation can be established,
\begin{equation}
    \log_{10}A_0^C = \alpha + \beta\log_{10}E_0\,.
    \label{eq:energy_cal}
\end{equation}

Putting all these relationships together, simulations such as those shown in Figs.\,\ref{fig:curves} and \ref{fig:calibration} can be used to estimate the energy of a cosmic-ray event recorded aboard a balloon-borne radio detector. The entire procedure is illustrated in Fig.\,\ref{fig:calibration} using a mock event, represented by a cross symbol.

First, the measured frequency spectrum for that event is fitted to Eq.\,\eqref{eq:expmodel}, and the spectral parameters $(A_0^{\rm sig},\,\gamma^{\rm sig})$ in some frequency range can be obtained. Once a shower geometry compatible with the observations is assumed (see Sections\,\ref{sec:direct_geom} and \ref{sec:reflex_geom}), simulations for several primary energies allow to obtain the dependence of the spectral amplitude $A_0$ with $\gamma$, as in Eq.\,\eqref{eq:cal_line} and as shown in the top panel of Fig.\,\ref{fig:calibration}. Such relations allow to obtain the spectral amplitude that would produce the event at the Cherenkov angle $A_0^{C,\,{\rm sig}}$ represented by a solid square symbol in Fig.\,\ref{fig:calibration}. Since this amplitude scales almost linearly with the logarithm of the primary energy as shown in the bottom panel of Fig.\,\ref{fig:calibration}. the value of $\log_{10}A_0^{C,\,{\rm sig}}$ yields an estimate of the energy of the mock event. 

This procedure can be applied to both direct and reflected events as described in Sections\,\ref{sec:direct} and \ref{sec:reflex} respectively.

\section{Energy reconstruction of direct cosmic ray events}\label{sec:direct}

During the first, third and fourth flights of the ANITA experiment, a total of 7 radio pulses compatible with atmosphere-skimming air showers were detected \cite{ANITAEnergyFlux, ANITAIII, ANITAIV}. Such particle cascades are attributed to cosmic rays inducing air showers developing entirely along the atmosphere with axis that do not intercept the Earth's surface. An example of such geometry is sketched in the bottom panel of Fig.\,\ref{fig:geometry}. At the time of their detection and the analysis of the ANITA I dataset \cite{ANITAEnergyFlux}, no simulation tools were available to model atmosphere-skimming air showers, preventing a detailed analysis of these events. In this Section, we introduce for the first time an energy reconstruction framework for atmosphere-skimming air showers.

These showers develop across low-density layers of the atmosphere leading to extremely elongated shower profiles reaching on the order of $\sim100\unit{km}$ length. The development across such long distances allows for a significant charge separation in the direction parallel to the Lorentz force, producing largely asymmetric shower fronts that are stretched along the direction of $\vect{v}\times\vect{B}$ \cite{RASPASS_Showers}, with $\vect{v}$ parallel to the shower axis and $\vect{B}$ the magnetic field of the Earth.

Both the increased dimensions of the air showers, as well as the asymmetry of the shower front, contribute towards increasing time delays between emissions from different regions of the shower development. This tends to reduce the high-frequency content of radio signals and thus generate steeper spectra compared to downward-going air showers.
Such behavior is visible in Fig.\,\ref{fig:pulses}, where for the same angular displacement from the Cherenkov angle, the high frequency components are more suppressed in atmosphere-skimming events than in downward-going showers.

Another distinctive feature of atmosphere-skimming events, is that the distance between the shower maximum and a balloon-borne detector can reach up to $\sim 900-1000\unit{km}$. The propagation of radio signals across such long distances through the gradient of atmospheric refractive index induces severe asymmetries in the radio emission pattern around the shower. Indeed, Cherenkov rings are in general not centered around the cascade axis and become elongated towards lower altitudes \cite{RASPASS_Radio}. As a consequence, the standard approximation $\psi_C\approx\arccos\left(1/n_X\right)$, with $n_X$ the refractive index at the position of shower maximum, is no longer valid in these events.

Due to this complex phenomenology, a realistic treatment of atmosphere-skimming air showers requires an accurate description of the shower geometry, magnetic field orientation, and radio-wave propagation that is event-specific. In the framework presented here, these effects are incorporated self-consistently through microscopic simulations performed for every recorded event \cite{RASPASS_Showers,RASPASS_Radio}. Setting up these simulations requires to specify the geometry of the shower axis, that as stated before cannot be determined just from the arrival direction of the signal at the payload.

To overcome this ambiguity, the event energy is estimated by averaging the energies reconstructed across multiple shower geometries compatible with the measured arrival direction. The associated uncertainty in the reconstructed energy is given by their dispersion. Identifying this set of compatible geometries is crucial for guiding the simulations required for calibration.

\subsection{Selection of compatible shower geometries}\label{sec:direct_geom}

The procedure to identify shower geometries compatible with an observed radio pulse is based on the knowledge of the payload position and the incoming direction of the pulse $(\theta,\phi)$ at the detector. This information is guaranteed for any balloon-borne radio array building up on the capabilities reached by ANITA. In the absence of prior information regarding the primary energy or particle type, the energy reconstruction must be based on simulations that represent average cosmic-ray showers at each primary energy, $E_0$. For radio emission, the two most relevant shower properties are the depth of shower maximum, $X_{\rm max}$, and the fraction of the primary energy transferred to the electromagnetic component of the shower, $f_{\rm EM}$, which determine the source-observer geometry and the strength of the radio emission, respectively.

The identification of compatible shower geometries begins by specifying an average depth of shower maximum, $\langle X_{\rm max}\rangle(E_0)$, obtained either from experimental measurements \cite{auger:xmaxphaseI} or from air-shower simulations. Because the bulk of the coherent radio emission is generated around the shower maximum, we approximate the observed radiation as originating only from $X_{\rm max}$. Consequently, the set of compatible shower geometries is constrained to those where $\langle X_{\rm max}\rangle$ lies along the reconstructed arrival direction of the radio pulse.

\begin{figure}
    \centering
    \includegraphics[width=\linewidth]{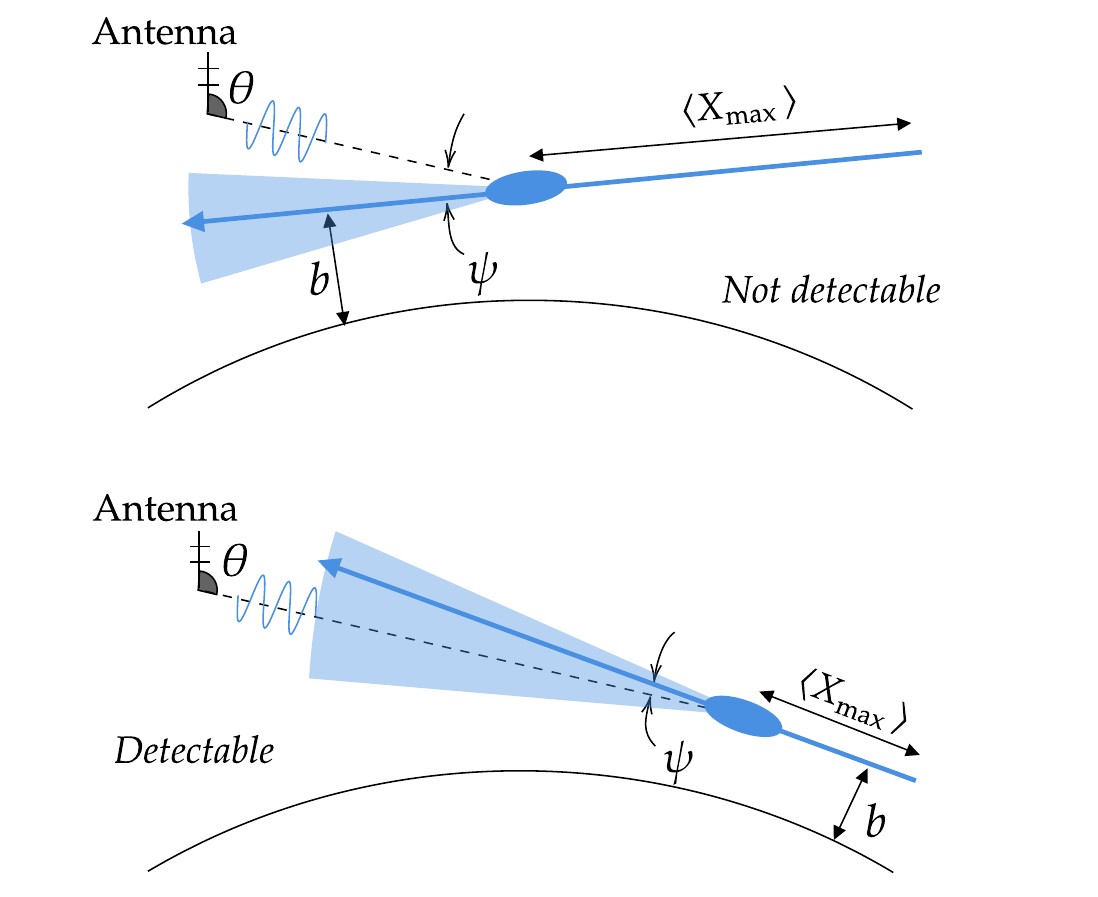}
    \caption{Sketch of the shower selection to performe calibration simulations in direct cosmic ray events. Two shower geometries (in blue) with different impact parameter $b$, are compatible with an observation of $X_{\rm max}$ with the same pulse direction $\theta$. The distance between shower maximum and observer, and the off-axis angle $\psi$ between the shower direction and the observer position, are different for each shower geometry. The Cherenkov cone is sketched in light blue. Both the difference between shower geometries compatible with the arrival of the pulse, and the opening angles of the Cherenkov cone, are largely exaggerated for illustration purposes.}
    \label{fig:showerselecdirect}
\end{figure}

By scanning along the arrival direction of the pulse, we identify several shower trajectories that reach $\langle X_{\rm max}\rangle$ at different spatial positions for each primary energy $E_0$ to be simulated. An illustrative example is shown in Fig.\,\ref{fig:showerselecdirect}, where two distinct shower trajectories (with different impact parameters) reach their shower maximum along the same radio-pulse incoming direction $\theta$.

The radio emission from air showers is typically observed near the Cherenkov angle, where the electric field is stronger. However, once the expected $\langle X_{\rm max}\rangle$ and the pulse arrival direction are fixed, not all geometric solutions will place the balloon payload inside the cone where emission is expected to be most intense. This is sketched in Fig.\,\ref{fig:showerselecdirect}. The shower in the top panel is only compatible with the pulse direction if the observer is located at a large off-axis angle $\psi > \psi_C$. Conversely, the geometry in the bottom panel allows the observer near the Cherenkov angle.

To exclude physically unlikely geometries, we require that the observer position can be illuminated by the Cherenkov cone. In the example of Fig.\,\ref{fig:showerselecdirect}, only the bottom trajectory would fulfill this requirement. Operationally, this is achieved by comparing the observer's off-axis angle $\psi$ with the Cherenkov angle $\psi_C$ for each candidate trajectory. Importantly, radiation pattern asymmetries caused by atmospheric refraction are naturally accounted for by numerically calculating $\psi_C$ at this step. Alternative criteria, allowing larger off-axis angles, could be employed if the measured spectral slope $\gamma$ is very steep.

Applying this criterion yields a set of \textit{compatible} shower geometries, ensuring that $\langle X_{\rm max}\rangle$ can be viewed close to the Cherenkov angle and precisely along the detected signal direction \footnote{Each \textit{compatible} trajectory can still be rotated azimuthally around the line of sight without changing the spatial position of $X_{\rm max}$. These rotated axes remain compatible with the observation but result in larger off-axis angles. Our method effectively samples this phase space by simulating observers at increasingly larger off-axis angles, up to $2\psi_C$. While this rotation slightly alters the relative orientation of the geomagnetic field (and thus the Lorentz force strength $\propto|\vect{v}\times\vect{B}|$), the effect on the radio emission is negligible since the angular uncertainty is tightly constrained by the Cherenkov cone ($\sim\psi_C \lesssim 1^\circ$).}. For each validated geometry, a preliminary batch of air showers is simulated. From this batch, we select the specific events whose $X_{\rm max}$ and electromagnetic energy fraction $f_{\rm EM}$ \footnote{Defined as the sum of the energy deposited in the atmosphere by $e^\pm$ and photons, plus the energy carried by particles falling below the simulation tracking threshold.} are closest to the sample averages.

From these sets of simulations, the dependence of the spectral parameters $A_0$ and $\gamma$ on the observer position and primary energy can be determined through calibration curves similar to those shown in Fig.\,\ref{fig:calibration}. These simulations are then directly used to reconstruct the energy of the event following the method described in Sec. \ref{sec:method}.

It is important to note that the extreme nature of atmosphere-skimming showers implies that the distance to $X_{\rm max}$ and the resulting radio emission pattern are highly sensitive to the specific shower geometry, the atmospheric density profile, and the local magnetic field. Because our methodology selects compatible trajectories using detailed atmospheric models and simulates the cascades within realistic IGRF magnetic fields \cite{igrf14}, these intricate systematic effects are naturally built into the energy reconstruction framework.

\subsection{Validation of the energy reconstruction method}\label{sec:direct_result}

In this Section we assess the performance of the energy reconstruction method described above, applied for the first time to \textit{direct} events. This is done by reconstructing the primary energy from a set of simulated radio pulses, produced by atmosphere-skimming air showers with different primary particles, energies and trajectories and observed at $36\unit{km}$ above sea level. The energies of these \textit{reference} events to be reconstructed were sampled from a uniform distribution in $\log_{10}\left(E_0/\unit{eV}\right)\in[17.5,20]$, with the incoming angle of the received pulses $\theta\in  [93.5^\circ,96^\circ]$. Showers with $\theta\lesssim 93^\circ$ were not simulated since they develop through very rarefied air, with a density around $4-5\%$ of the sea level value. For those geometries, the shower maximum happens close to the position of a detector at $36\unit{km}$ above sea level or even beyond it, potentially affecting the coherence of the emission due to close-source effects. On the other hand showers with $\theta>96^\circ$ (the angle at which the horizon is seen from an altitude of $36\unit{km}$) are not atmosphere-skimming. No cut on the observer angular position $\psi$ was used to select the reference events, since the Cherenkov angle $\psi_C$ is not uniquely defined for atmosphere-skimming showers. Instead, we required a  signal spectrum unaffected by numerical noise in the simulations at least up to $400\unit{MHz}$. The simulations were performed with \textsc{ZHAireS-RASPASS}, using a vertical magnetic field of intensity $50\unit{\mu T}$. 

The \textit{true} position of $X_{\rm max}$ was used to find the \textit{true} incoming direction $\theta$ of the pulse for each reference event to be reconstructed. The value of $\theta$ was subsequently adopted to build the calibration simulations needed for the energy reconstruction as explained in Sec.\,\ref{sec:direct_geom}.

The performance of the energy estimation method depends on the uncertainty on the spectral parameters $A_0$ and $\gamma$ obtained from the measured signal, and on the ambiguity in the shower axis direction. The unknown nature of the primary particle introduces also an additional source of uncertainty. In order to decouple these effects, we performed a first validation of the energy reconstruction  assuming a perfect knowledge of the incoming direction $\theta$ and the primary composition (in this case 100\% protons), and in the absence of any environmental noise. Under these ideal conditions, the performance of the energy estimation method is mainly driven by the capability to simulate showers with geometries compatible with the true event direction.

For each reference event, we found five different sets of compatible shower geometries, each set containing simulations at several primary energies. These showers were used to produce calibration plots similar to those in Fig.\,\ref{fig:calibration}, up to maximum off-axis positions of $2\psi_C$. The choice of shower geometries was done by targeting the average $\langle X_{\rm max}\rangle$ depth for proton showers as predicted by \textsc{ZHAireS} with the same hadronic model used in simulations, in our case \textsc{Sibyll 2.3d} \cite{Sibyll23d}. 

\begin{figure*}
    \centering
    \includegraphics[width=.9\linewidth]{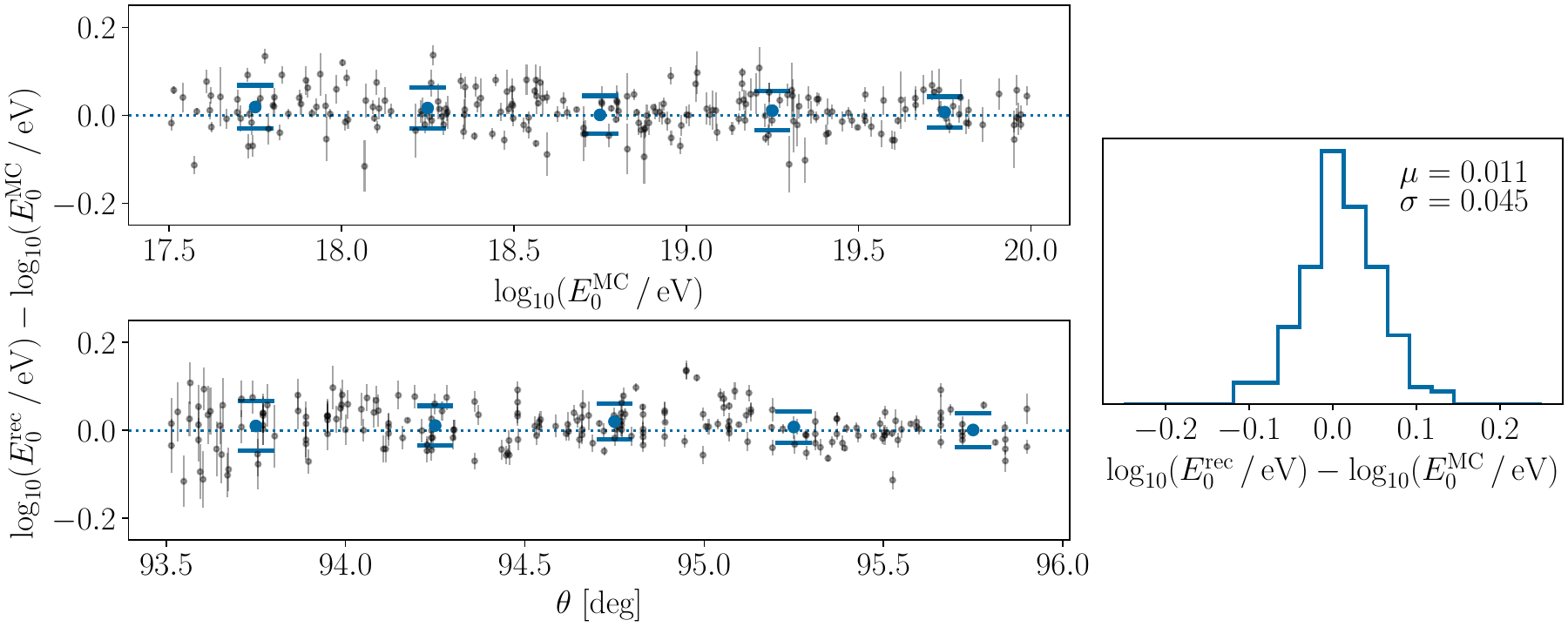}
    \caption{Results of the energy reconstruction method in the $100-450\unit{MHz}$ frequency range, applied to 250 simulated radio pulses produced by atmosphere-skimming proton air showers of several primary energies and impact parameters. Top left panel: Bias in the reconstructed $\log_{10}\left(E_0^{\rm rec}\,/\unit{eV}\right)$ as a function of the Monte Carlo energy $\log_{10}\left(E_0^{\rm MC}\,/\unit{eV}\right)$. The average and standard deviation of the reconstruction bias, in bins of width $0.5$ in $\log_{10}\left(E_0\,/\unit{eV}\right)$, are shown with blue markers. Bottom left panel: Bias in the reconstructed $\log_{10}\left(E_0^{\rm rec}\,/\unit{eV}\right)$ as a function of the incoming angle of radiation $\theta$ seen at the position of the detector. The average and standard deviation of the reconstruction bias, in bins of width $0.5^\circ$, are overlaid with blue markers. Right panel: Distribution of biases in the reconstructed $\log_{10}\left(E_0\,/\unit{eV}\right)$ for all events.}
    \label{fig:reco_direct_nobias}
\end{figure*}

In this first evaluation of the method, we obtained spectral parameters using the amplitude spectrum in the $100-450\unit{MHz}$ frequency band. This range is contained in the bandwidth of the Low-Frequency (LF) instrument of the PUEO experiment \cite{PUEO_LF_ARENA24}, specifically designed to increase the sensitivity of this detector to air-shower events. The results of the energy reconstruction method are shown in Fig.\,\ref{fig:reco_direct_nobias}, where the reconstructed energy, $E_0^{\rm rec}$, is compared to the true simulated energy, $E_0^{\rm MC}$.

The true energy of the events is recovered accurately on average, without evident biases as a function of the primary energy or the shower geometry. The spread of the distribution around the true $\log_{10}\left(E_0/\unit{eV}\right)$ corresponds to a $\sigma(E)/E\sim11\%$ resolution in the primary energy. As mentioned before, this spread is dominated by the intrinsic uncertainty in the shower axis orientation, as well as event-to-event fluctuations around the average values of $X_{\rm max}$ and $f_{\rm EM}$ at each energy. Indeed, the slight asymmetry in the distribution of reconstructed $\log_{10}\left(E_0^{\rm rec}/\unit{eV}\right)$ around the true values is dominated by those reference events with $X_{\rm max}$ significantly deeper than the expected average. In these events, both the increased electromagnetic energy fraction (that correlates with deeper $X_{\rm max}$ \cite{RASPASS_Showers}) and the fact that the showers are closer to the detector, result in a slight and expected overestimation of the true energy. Still, the resolution in the reconstructed energy is comparable with the previous results of \cite{ANITAEnergyFlux}.

Fluctuations of just a few $\unit{g/cm^2}$ around the average $X_{\rm max}$ can significantly change the position of shower maximum in atmosphere-skimming showers. These event-to-event fluctuations are responsible for the worse resolution in the reconstructed energy as the incoming angle of the pulse $\theta$ decreases (bottom panel of Fig.\,\ref{fig:reco_direct_nobias}) approaching $\theta=93^\circ$. For the most horizontal pulses, the impact of fluctuations in the position  of $X_{\rm max}$ in $\unit{g/cm^2}$ translate into larger fluctuations in distance due to the lower densities where the shower development takes place.

\subsubsection{Detector pointing resolution}\label{sec:direct_angularbias}

An accurate knowledge of the incoming direction $\theta$ of the radio signal is crucial in the energy reconstruction method, in particular for the selection of shower geometries compatible with the detected pulse. 
Balloon-borne radio payloads typically use interferometric techniques to obtain the direction of the signal, by comparing the arrival time delays of the pulses across different antennas \cite{Romero-Wolf:2014pua}. The pointing resolution of such techniques is intrinsically limited by both the size of the payload and its frequency range of operation, as well as by the timing accuracy between individual antennas. 

Given the nearly horizontal trajectories of atmosphere-skimming events, a minor change in the shower axis drastically alters the atmospheric density profile traversed by the cascade, which in turn significantly shifts the position of $X_{\rm max}$. As an example, for pulses with incoming directions $\theta=94.5^\circ$ and $\theta = 95^\circ$ the  difference in the expected position of $X_{\rm max}$ is $\sim 120\unit{km}$  \footnote{ In this case the distance between the expected $X_{\rm max}$ and a balloon-borne payload at an altitude of $36\unit{km}$ ranges from $\sim780\unit{km}$ to $\sim900\unit{km}$.}.

\begin{figure*}
    \centering
    \includegraphics[width=.9\linewidth]{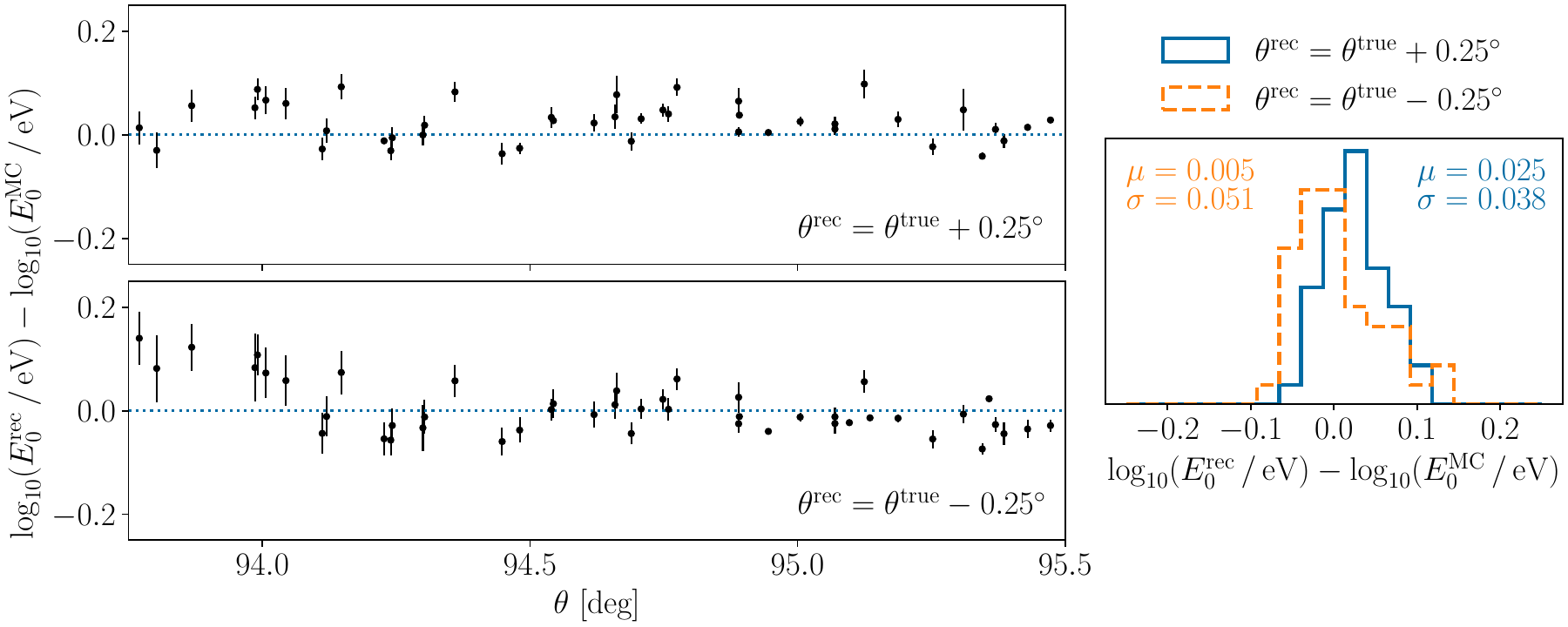}
    \caption{Results of the energy reconstruction method in the $100-450\unit{MHz}$ frequency range, applied to a subset of 50 simulated radio pulses produced by atmosphere-skimming proton air showers of several primary energies and impact parameters. Left panels: Bias in the reconstructed $\log_{10}\left(E_0^{\rm rec}\,/\unit{eV}\right)$ as a function of the incoming angle of radiation $\theta$ at the position of the detector. The incoming angle assumed in the reconstruction of the energy of the events, $\theta^{\rm rec}$, is deliberately $0.25^\circ$ larger (smaller) than the true value $\theta^{\rm true}$ in the top (bottom) panel. Right panel: Distribution of biases in the reconstructed $\log_{10}\left(E_0\,/\unit{eV}\right)$ for all events, in the two scenarios $\theta^{\rm rec}=\theta^{\rm true}\pm0.25^\circ$.}
    \label{fig:reco_direct_angularbias}
\end{figure*}

In order to evaluate the impact of a finite pointing resolution, we applied the energy reconstruction method to a subset of 50 simulated radio signals produced by atmosphere-skimming proton air showers, deliberately introducing a constant bias of $\pm0.25^\circ$ in the incoming angle $\theta$ of the radiation with respect to the true value, comparable to the pointing resolution of the ANITA experiment \cite{ANITAIII, ANITAIV}. The method was again applied assuming a perfect knowledge of the primary mass composition and in the absence of noise. The results are shown in Fig.\,\ref{fig:reco_direct_angularbias}. 

Building the calibration simulations with an assumed incoming angle larger than the true value ($\theta^{\rm rec} > \theta^{\rm true}$) selects showers that develop in denser regions of the atmosphere. Because this places $X_{\rm max}$ farther away from the detector, a higher primary energy is required in the simulation to reproduce the observed signal amplitude. Consequently, the reconstructed energy is slightly overestimated, as shown in the top panel of Fig.\,\ref{fig:reco_direct_angularbias}.

Conversely, when the assumed angle is smaller than the true value ($\theta^{\rm rec} < \theta^{\rm true}$), the reconstructed energy is generally underestimated, as seen in the bottom panel of Fig.\,\ref{fig:reco_direct_angularbias} for $\theta^{\rm true} \gtrsim 94^\circ$. However, for events with $\theta^{\rm true} \lesssim 94^\circ$ this behavior inverts and the energy is once again overestimated. This inversion occurs because, in this angular regime, reducing the incoming angle corresponds to selecting nearly horizontal showers that develop much higher in the atmosphere. Even though $X_{\rm max}$ is placed closer to the detector, these high-altitude showers are significantly more elongated \cite{RASPASS_Showers}. This extreme longitudinal extent reduces the coherence of the simulated radio emission \cite{RASPASS_Radio}, forcing the reconstruction algorithm to compensate with a higher primary energy to match the observed signal amplitude.

\subsubsection{Primary mass assumption}\label{sec:direct_massbias}

The parameterizations $\langle X_{\rm max}\rangle(E_0)$, that are needed to select shower geometries compatible with the observed pulses, as well as the fraction of primary energy $f_{\rm EM}$ contributing to the radio emission, depend on the assumed primary mass. Proton-induced showers have deeper $X_{\rm max}$ and larger electromagnetic energy fractions than iron showers with the same geometry and primary energy, leading to larger signal amplitudes at the detector.  

\begin{figure*}
    \centering
    \includegraphics[width=.9\linewidth]{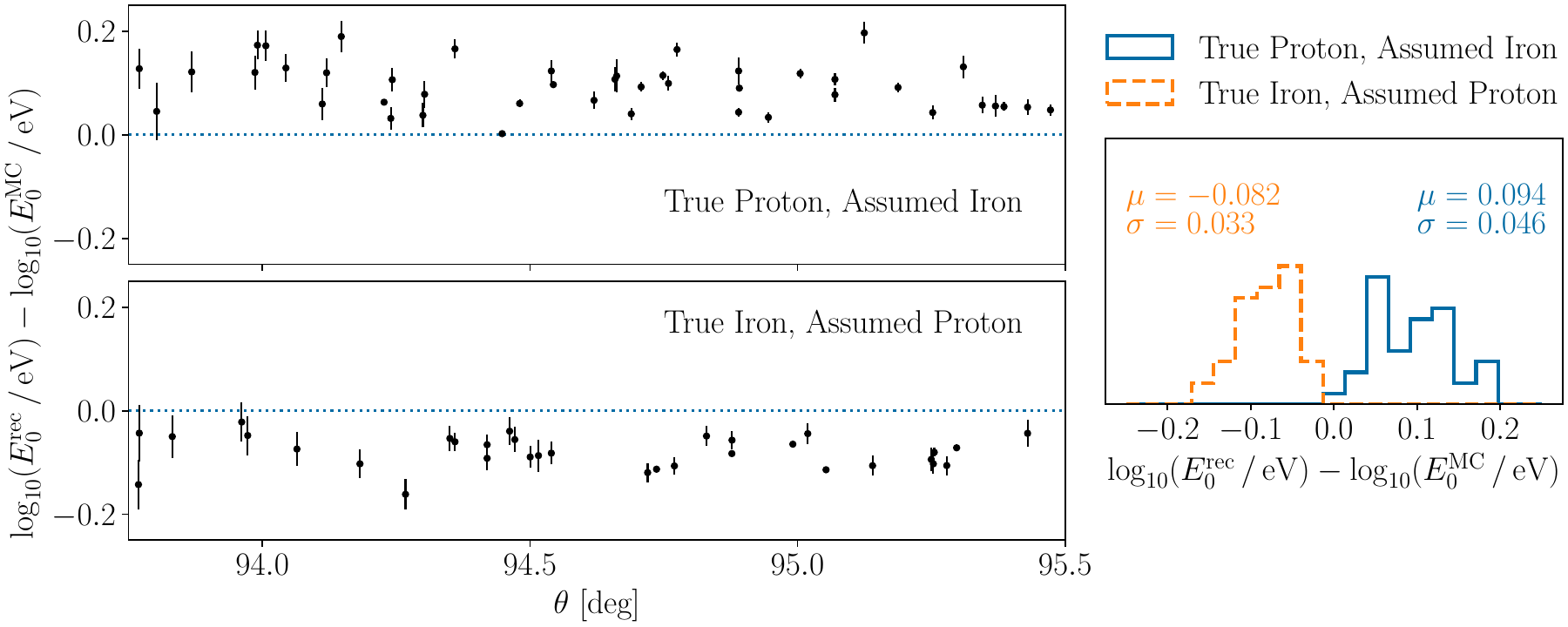}
    \caption{Results of the energy reconstruction method in the $100-450\unit{MHz}$ frequency range, applied to a subset of 50 simulated radio pulses produced by atmosphere-skimming proton and iron air showers of several primary energies and impact parameters. Top left panel: Bias in the reconstructed $\log_{10}\left(E_0^{\rm rec}\,/\unit{eV}\right)$ as a function of the incoming angle of radiation $\theta$ for proton events reconstructed under the assumption of a primary iron. Bottom left panel: Same as the top panel, for iron events reconstructed under the assumption of a primary proton. Right panel: Distribution of biases in the reconstructed $\log_{10}\left(E_0\,/\unit{eV}\right)$ for all events in the two scenarios of primary mass misidentification.}
    \label{fig:reco_direct_massbias}
\end{figure*}

Since in a real event the primary mass is unknown, we evaluated the impact of the primary mass assumption on the energy reconstruction using a worst-case scenario. We applied the energy reconstruction method to a set of radio signals produced by 50 proton air showers, but adopting the  $\langle X_{\rm max}\rangle(E_0)$ of iron primaries when selecting compatible shower geometries; and vice-versa. The energy reconstruction is performed under a perfect knowledge of the pulse incoming direction $\theta$ and in the absence of noise to isolate the effect of the unknown primary mass. The results obtained under these conditions are shown in Fig.\,\ref{fig:reco_direct_massbias}.

As expected, a clear bias in the reconstructed energies appears. In the case of proton-air showers reconstructed under the assumption of an iron primary, the \textit{true} events have on average larger electromagnetic energy fractions and deeper (i.e. closer to the detector) $X_{\rm max}$ than the simulations used to reconstruct the energy. As a consequence, the energy needed by an iron primary to reproduce the observed signal must be larger. This effect is inverted in the case of iron-induced signals reconstructed under the assumption of a primary proton. The biases of $\mu_{\log\left(E\right)}\sim\pm0.09$ correspond to a $\sim25\%$ over- or underestimation of the primary energies in this worst-case scenario, with maximal difference between primary masses. 

The application of this method to actual events should thus be made including an adequate definition of the average cosmic ray event at each energy, for instance targeting events with the average $X_{\rm max}$ reported by high-exposure cosmic-ray experiments and including the measured evolution of the average primary mass as a function of the energy \cite{PierreAuger:2025rdo}.

\subsubsection{Frequency range and environmental noise}\label{sec:direct_noisebias}

In a real detector, the performance of the energy reconstruction method will also be affected by the accuracy with which the spectral parameters $A_0$ and $\gamma$ can be recovered from a measurement in the presence of noise and the frequency range where the spectral fits are performed.

The effects of the noise level and frequency band in the energy estimation method were evaluated with the same sample of 250 events used in the first validation of the method (Fig.\,\ref{fig:reco_direct_nobias}), for different Signal-to-Noise Ratios (SNR) and performing the spectral fits in two different frequency ranges: $100-450\unit{MHz}$, matching the bandwidth of the LF instrument of PUEO and PBR; and $300-1000\unit{MHz}$, the same frequency range used in the analysis of the ANITA I events and contained within the bandwidth of the Main Instrument (MI) of PUEO. We assume in this case a perfect pointing resolution and complete knowledge of the primary composition. 

To emulate realistic observing conditions, the 250 clean radio pulses were injected with Gaussian noise of a fixed RMS, and then processed with a passband filter in the $50-1200\unit{MHz}$ frequency range to match the combined bandwidth of the PUEO instrument. The RMS of the injected noise was chosen to yield specific, fixed SNR levels:

\begin{equation}
    \text{SNR}=\frac{\mathcal{E}_{\rm max} - \mathcal{E}_{\rm min}}{2\sqrt{\langle \mathcal{N}^2\rangle}},
    \label{eq:SNR}
\end{equation}

\noindent where $\mathcal{E}(t)$ and $\mathcal{N}(t)$ represent the time traces of the signal electric field and noise in the passband, respectively. The treatment of these \textit{reference} signals follows a similar approach to that of \cite{ANITAEnergyFlux, Martinelli_RiceMethod}. The time trace is splitted into a \textit{signal} window, containing the main signal peak, and ten \textit{noise} windows where the average noise spectrum is estimated. After that, maximum-likelihood estimators for the underlying signal are obtained at each frequency bin in the signal window, assuming a Rice distribution for the spectral amplitude in presence of the measured noise spectrum. In this particular case, we used windows of width $30\unit{ns}$, amounting to a total trace length of $330\unit{ns}$ sampled at $10\unit{GSa/s}$. 

\begin{figure}
    \centering
    \includegraphics[width=\linewidth]{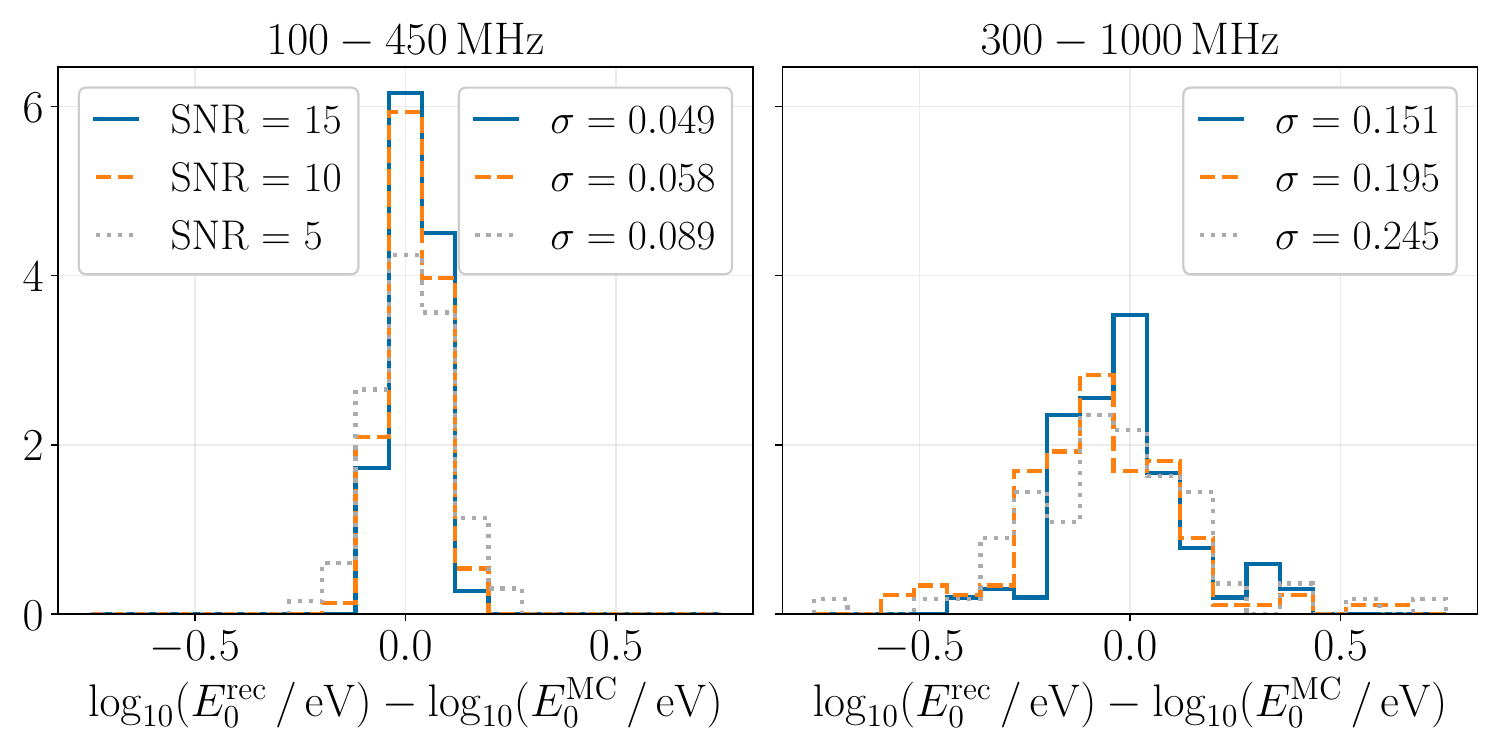}
    \caption{Left panel: Normalized distribution of biases in the reconstructed $\log_{10}\left(E_0\,/\unit{eV}\right)$ for atmospheric-skimming events, after performing the spectral fits in the $100-450\unit{MHz}$ frequency range, under various SNR levels. Right panel: Same as left panel, performing the spectral fits in the $300-1000\unit{MHz}$ frequency range. The standard deviation $\sigma\left(\log_{10}E\right)$ is indicated in both panels.}
    \label{fig:reco_direct_noisebias}
\end{figure}

The results of the energy reconstruction method, evaluated in the presence of noise and across the frequency bands of interest, are shown in Fig.\,\ref{fig:reco_direct_noisebias}. The resolution in the reconstructed energies degrades significantly when higher-frequency components of the signal are included in the $300-1000\unit{MHz}$ band. This is expected because higher frequencies are more sensitive to time delays between emissions from different regions of the shower and hence to event-to-event fluctuations, especially in the position of $X_{\rm max}$. Furthermore, due to the exponentially decreasing shape of the signal spectra, high-frequency components are more susceptible to fluctuations in the noise floor. As expected, lowering the SNR results in an moderate degradation of the resolution in the reconstructed energies in both frequency ranges. Nevertheless, the absence of any significant bias in the estimated $E^{\rm rec}_0$ demonstrates that an accurate energy reconstruction is fully achievable in the presence of broadband background noise.

\section{Energy reconstruction of reflected cosmic-ray events}\label{sec:reflex}

The energy estimation method described in Sec. \ref{sec:method} was originally applied in \cite{ANITAEnergyFlux} to a set of 14 reflected radio pulses recorded by the first flight of the ANITA experiment. This resulted in the first measurement of the cosmic-ray energy spectrum using the radio technique alone, yielding a result in agreement with the spectrum measured with the Pierre Auger Observatory \cite{PierreAuger:2025eun}. 

In this Section we provide a major update of the energy reconstruction method for reflected events, including modifications of the treatment of the signal reflection off a spherical surface that have been implemented directly in the \textsc{ZHAireS} simulation framework \cite{Reflex}.

A proper description of the reflecting surface is crucial to understand the properties of reflected radio pulses \cite{Gorham:2017xbo, Prohira:2018mmv, ANITAEnergyFlux}. At the time of the analysis of the ANITA I events, the simulations of radio pulses with \textsc{ZHAireS} treated reflections on the Antarctica modeling the surface as a perfectly smooth and flat dielectric surface with refractive index $n=1.35$ \cite{Reflex}. 

Furthermore, the decrease of signal amplitude caused by the divergence of rays after reflection on the spherical surface of the Earth, as well as the frequency dependent corrections due to surface roughness, were parametrized and applied to the simulated radio pulses as a whole before using them to estimate the energy of the recorded events \cite{ANITAEnergyFlux, Reflex}.

In this work, we have updated \textsc{ZHAireS} to consider reflections on a smooth spherical surface for every single particle track in the simulation. An iterative procedure is employed to find the position of the reflection point on a sphere, as well as the incidence angle of the radiation on the surface $\alpha$, so that the reflected ray connects the track to the payload. At the interface, the Fresnel coefficients are applied to each polarization of the electric field. In addition, a \textit{defocusing} term, representing the decrease in amplitude due to ray divergence after reflection on a sphere, is now included and applied to the emission of each particle track. Further details about this procedure can be found in Appendix \ref{app:reflex_mod}. 

Introducing reflections on a spherical surface enables a more accurate modeling of the reflected radio emission from highly-inclined downward-going air showers with zenith angle $\vartheta_S\gtrsim 80^\circ$. In these events, the projection of the Cherenkov ring on the ground can span several hundred kilometers on the curved surface of the Earth. As shown in Appendix \ref{app:reflex_mod}, this can lead to a significant widening of the reflected Cherenkov ring with respect to a reflection on a flat surface due to the divergence of the rays.

Accounting for reflections from a rough surface on a particle-by-particle basis would require a frequency-dependent correction to the emission of each track, with the impact of the roughness expected to be larger for the smaller wavelengths capable of resolving the details of the surface. This would require a significant modification of the time-domain implementation of the electric field in \textsc{ZHAireS} resulting in a significantly increased CPU time. For simplicity, we instead apply the same \textit{a posteriori} roughness correction as in \cite{ANITAEnergyFlux} to the simulated electric fields, maintaining consistency with the ANITA I analysis.

\subsection{Selection of compatible shower geometries}\label{sec:reflex_geom}

The selection of shower geometries compatible with the observed direction of the radio pulse proceeds similarly to that described in Sec. \ref{sec:direct_geom}. The relevant geometric constraints are the payload position and the incoming direction of the radio pulse $(\theta,\phi)$ at the detector. Together, these two elements allow us to determine the reflection point on the surface and the incidence angle $\alpha$ of the radiation on the ground. 

By assuming an average depth of shower maximum $\langle X_{\rm max}\rangle(E_0)$, we can scan the possible positions of shower maximum along the reflected pulse direction, and identify the shower geometries compatible with the event for each simulated primary energy $E_0$.

\begin{figure}
    \centering
    \includegraphics[width=\linewidth]{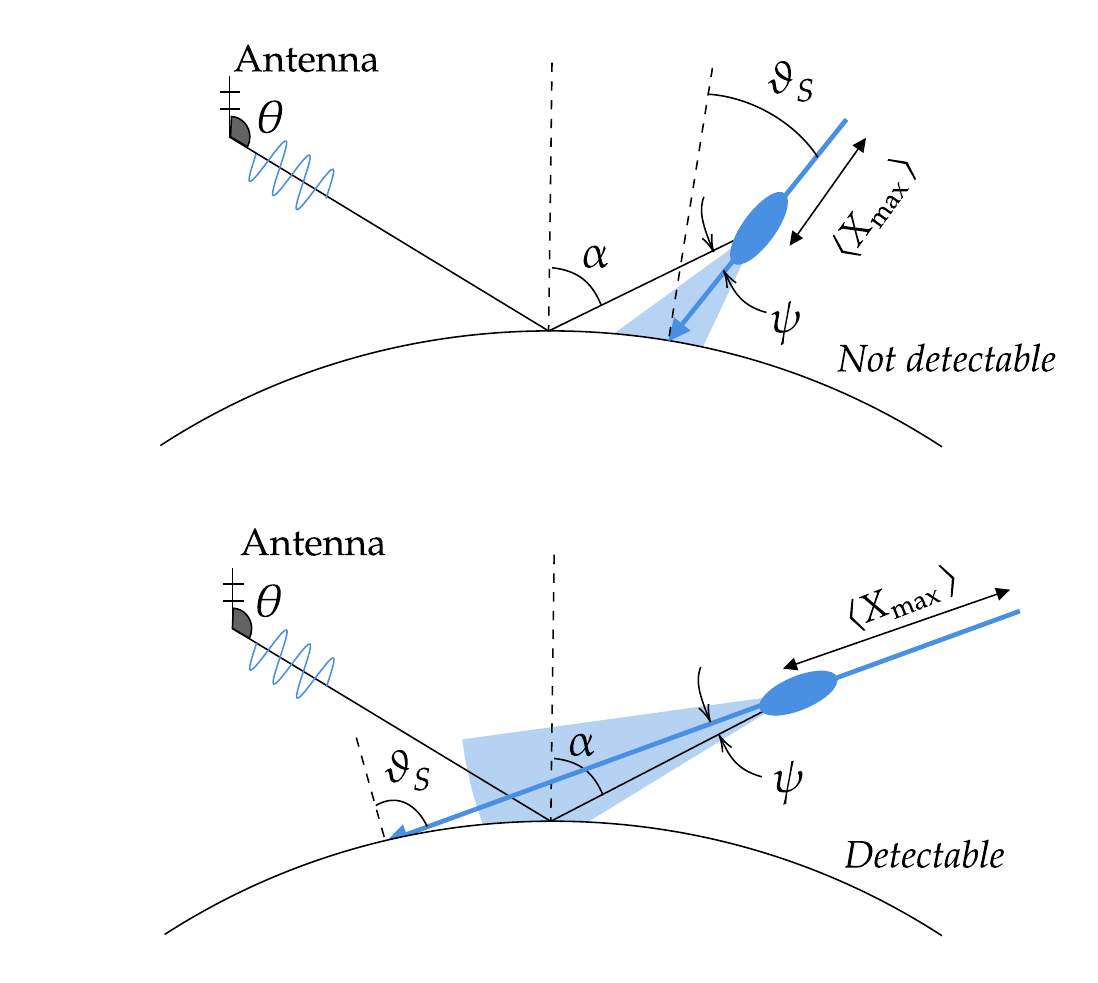}
    \caption{Sketch illustrating the selection of shower geometries compatible with the observed direction of the radio pulse in reflected cosmic ray events. Blue arrows indicate two shower geometries with the same value of $\langle X_{\rm max}\rangle$ but different zenith angle $\vartheta_S$. Shower maxima occur at different altitudes, due to the different $\vartheta_S$. Both shower geometries are compatible with the same pulse direction $\theta$ and incidence angle on ground $\alpha$. However, the off-axis angle $\psi$ between the shower direction and the observer position is different in each case. The Cherenkov cone is sketched in light blue. Both the difference between shower geometries compatible with the arrival of the pulse, and the opening angles of the Cherenkov cones, are largely exaggerated for illustration purposes.}
    \label{fig:showerselecreflex}
\end{figure}

In Fig.\,\ref{fig:showerselecreflex} we show two shower geometries with the same value of $\langle X_{\rm max} \rangle$ compatible with the direction of the pulse, but with their maxima at different altitudes and distances to the observer. As was the case for direct events, a further selection of shower geometries is made requiring that the reflection point on the ground can be illuminated by the Cherenkov cone. This guarantees that the simulated showers are observed close to the Cherenkov angle $\psi_C$ \footnote{For downward-going showers, the approximation $\psi_C\sim\arccos(1/n_X)$ with $n_X$ the refractive index at the altitude of $X_{\rm max}$ is adequate, and is used at this step.}. This condition is sketched in Fig.\,\ref{fig:showerselecdirect} where only the shower in the bottom panel would be selected for calibration. 

For each validated geometry, a preliminary batch of air showers is simulated. From this sample, the events whose $X_{\rm max}$ and $f_{\rm EM}$ are closest to the expected average values are selected. These average simulations are then used to determine the dependence of the spectral parameters $A_0$ and $\gamma$ on the off-axis angle $\psi$ and primary energy $E_0$, yielding calibration curves similar to those shown previously in Fig.\,\ref{fig:calibration}.

\subsection{Validation of the energy reconstruction method}\label{sec:reflex_result}

We now evaluate the performance of the energy reconstruction method for the case of \textit{reflected} events. The methodology follows the procedure of Sections \ref{sec:method} and \ref{sec:reflex_geom}. In this case, 250 \textit{reference} events to be reconstructed were sampled from a uniform distribution in $\log_{10}\left(E_0/\unit{eV}\right)\in [17.5,20]$, with pulses sampled in the incidence angle on ground $\alpha\in[65^\circ,90^\circ]$, and in the off-axis angle $\psi$ of the detector position measured with respect to $X_{\rm max}$, with $\psi\in[\psi_C-0.2^\circ,\psi_C+0.2^\circ]$. These events were simulated with \textsc{ZHAireS} including the reflection on a spherical layer of ice ($n=1.35$) placed at $2.5\unit{km}$ above sea level (see Appendix \ref{app:reflex_mod}), and adopting a vertical magnetic field of intensity $50\unit{\mu T}$. 

The incidence angle $\alpha$ was calculated using the true position of $X_{\rm max}$ in each reference event, and then used to build the compatible simulations needed for the energy reconstruction as explained above. 

A first validation of the energy estimation method under the assumption of perfect pointing resolution, knowledge of the primary mass and in the absence of noise, is shown in Fig.\,\ref{fig:reco_reflex_noisebias} for the $100-450\unit{MHz}$ frequency band.

\begin{figure*}
    \centering
    \includegraphics[width=.9\linewidth]{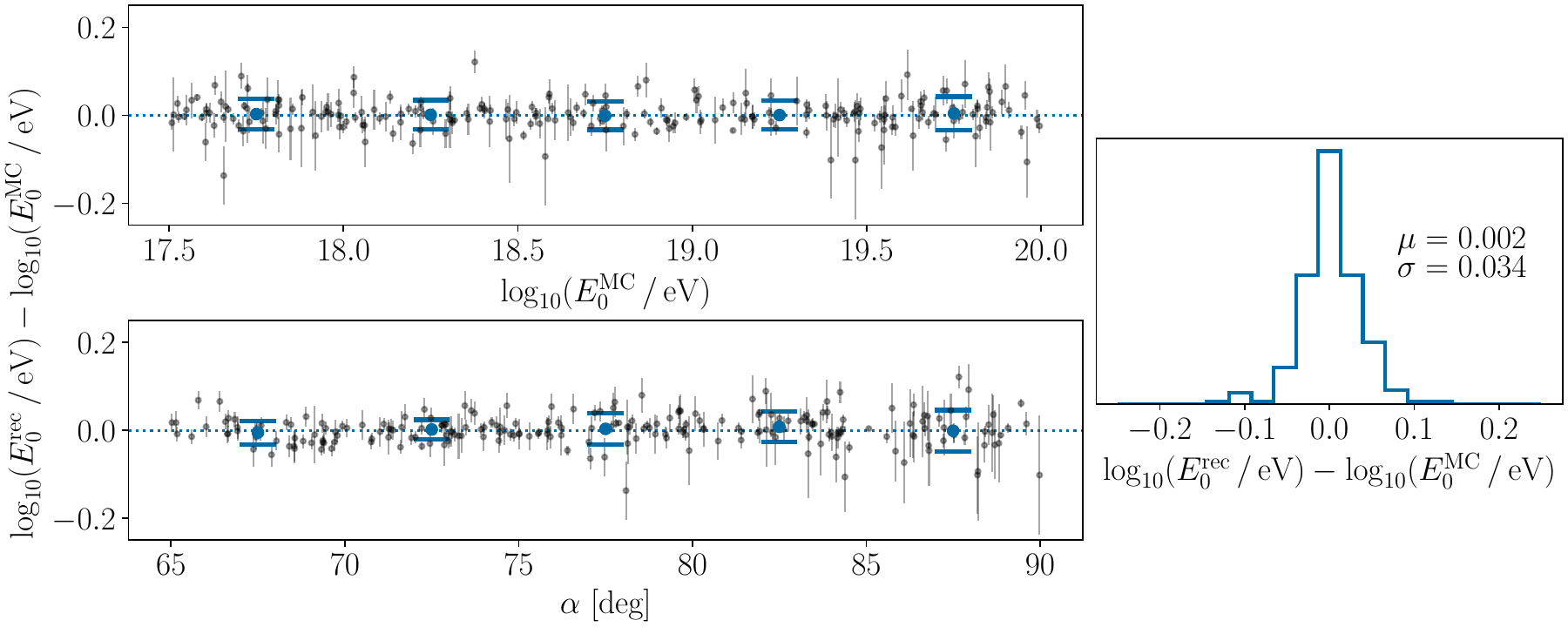}
    \caption{Results of the energy reconstruction method in the $100-450\unit{MHz}$ frequency range, applied to 250 simulated radio pulses produced by downward-going proton air showers of several primary energies and incidence angle of radiation on the ice. Top left panel: Bias in the reconstructed $\log_{10}\left(E_0^{\rm rec}/\unit{eV}\right)$ as a function of the Monte Carlo primary energy $\log_{10}\left(E_0^{\rm MC}/\unit{eV}\right)$. The mean and standard deviation of the reconstruction bias, in bins of width $0.5$ in $\log_{10}\left(E_0\,/\unit{eV}\right)$, are shown with blue markers. Bottom left panel: Bias in the reconstructed $\log_{10}\left(E_0^{\rm rec}/\unit{eV}\right)$ as a function of the true incidence angle of radiation on the ice $\alpha$. The mean and standard deviation of the reconstruction bias, in bins of width $5^\circ$, are overlaid with blue markers. Right panel: Distribution of  $\log_{10}\left(E_0^{\rm rec}/\unit{eV}\right)-\log_{10}\left(E_0^{\rm MC}/\unit{eV}\right)$ for the 250 reconstruced events.}
    \label{fig:reco_reflex_nobias}
\end{figure*}

On average, the energy reconstruction method is able to recover the true energy of the primary particle with a negligible bias in $\log_{10}(E)$ independently of primary energy and incidence angle. The spread of the distribution around the true energy, $\sigma_{\log\left(E\right)}\sim0.034$, corresponds to $\sigma(E)/E\sim8\%$ resolution in the primary energy. This is comparable to the result obtained for direct events, although slightly better because in these events the shower geometry is more tightly constrained and the effect of event-to-event fluctuations is less pronounced.

On the other hand, the resolution shows a tendency to decrease with $\alpha$, as can be seen in the bottom left panel of Fig.\,\ref{fig:reco_reflex_nobias}. 
Larger values of $\alpha$ correspond to more inclined showers (larger $\vartheta_S$), which develop in regions of lower atmospheric density, enhancing the effect of fluctuations in the depth of $X_{\rm max}$ of a few tens of $\unit{g/cm^2}$ on the distance to the detector.  

These results represent a best-case scenario, obtained under a perfect knowledge of the radiation incidence angle and primary cosmic ray mass, and in the absence of noise. The impact on the reconstruction of each of these variables is studied separately in the following.

\subsubsection{Detector pointing resolution}\label{sec:reflex_angularbias}

\begin{figure*}
    \centering
    \includegraphics[width=.9\linewidth]{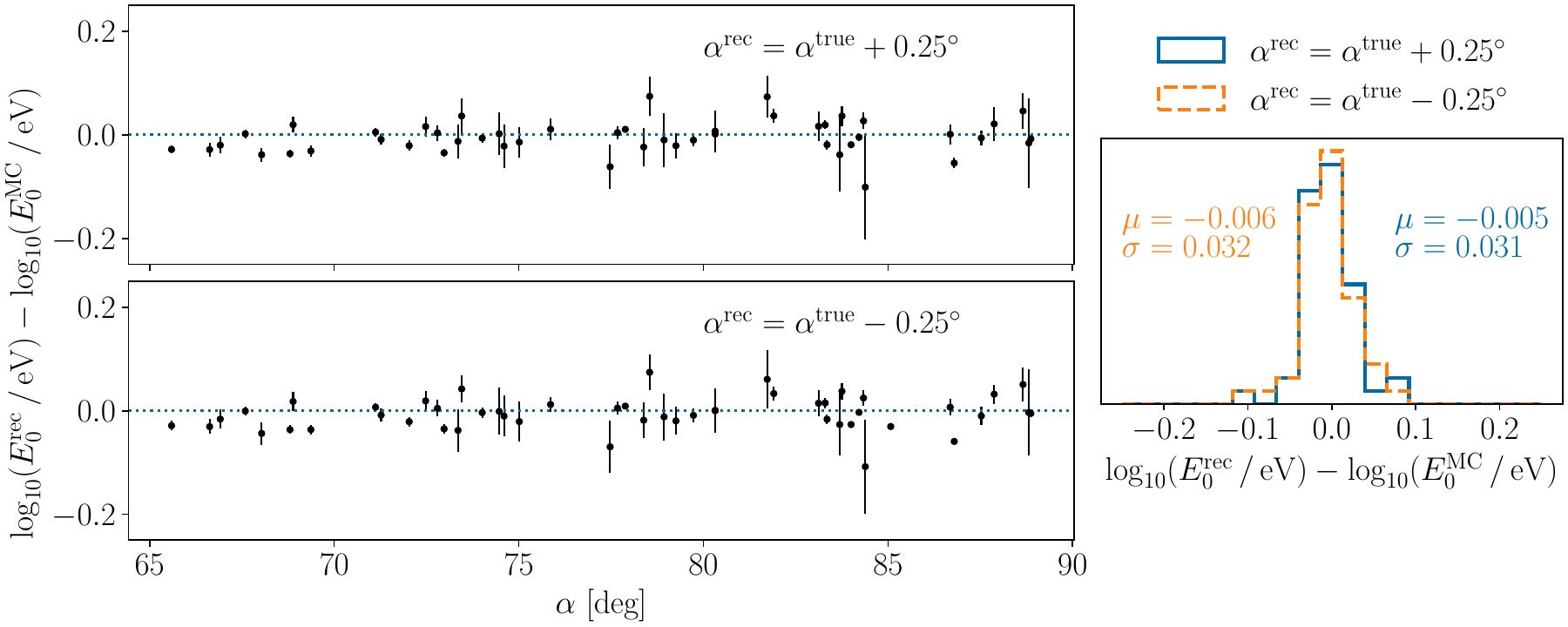}
    \caption{Results of the energy reconstruction method in the $100-450\unit{MHz}$ frequency range, applied to a subset of 50 simulated radio pulses produced by downward-going proton air showers of various energies and incidence angle of radiation on the ice. Left panels: Difference between the reconstructed $\log_{10}\left(E_0^{\rm rec}/\unit{eV}\right)$ and the true Monte Carlo energy $\log_{10}\left(E_0^{\rm MC}/\unit{eV}\right)$ as a function of the true incidence angle of radiation on the ice $\alpha$. The incidence angle used to reconstruct the energy of the events, $\alpha^{\rm rec}$, is deliberately $0.25^\circ$ larger (smaller) than the true value $\alpha^{\rm true}$ in the top (bottom) panel. Right panel: Distribution of differences for all events, in the two cases $\alpha^{\rm rec}=\alpha^{\rm true}\pm0.25^\circ$.}
    \label{fig:reco_reflex_angularbias}
\end{figure*}

In order to estimate the effects of a finite pointing resolution, we followed the same approach as in Sec. \ref{sec:direct_angularbias} and reconstructed the primary energies of a subset of 50 reference events, including a constant bias of $\pm0.25^\circ$ in values of the incidence angles needed to build the calibration simulations. The results of the energy reconstruction in this scenario, in absence of noise and with perfect knowledge of the primary particle, are shown in Fig.\,\ref{fig:reco_reflex_angularbias}. Despite the bias in $\alpha$, no clear shift in the reconstructed energies is appreciated.

For downward-going air showers, the properties of downward-going cascades used for calibration, such as the position of $X_{\rm max}$, or the corrections on the pulses due to reflection, do not change significantly when the shower geometry is modified by a fraction of a degree, and the impact of a bias of $\sigma_\alpha\sim0.25^\circ$ has a negligible effect for reflected events.

\subsubsection{Primary mass assumption}\label{sec:reflex_massbias}

\begin{figure*}
    \centering
    \includegraphics[width=.9\linewidth]{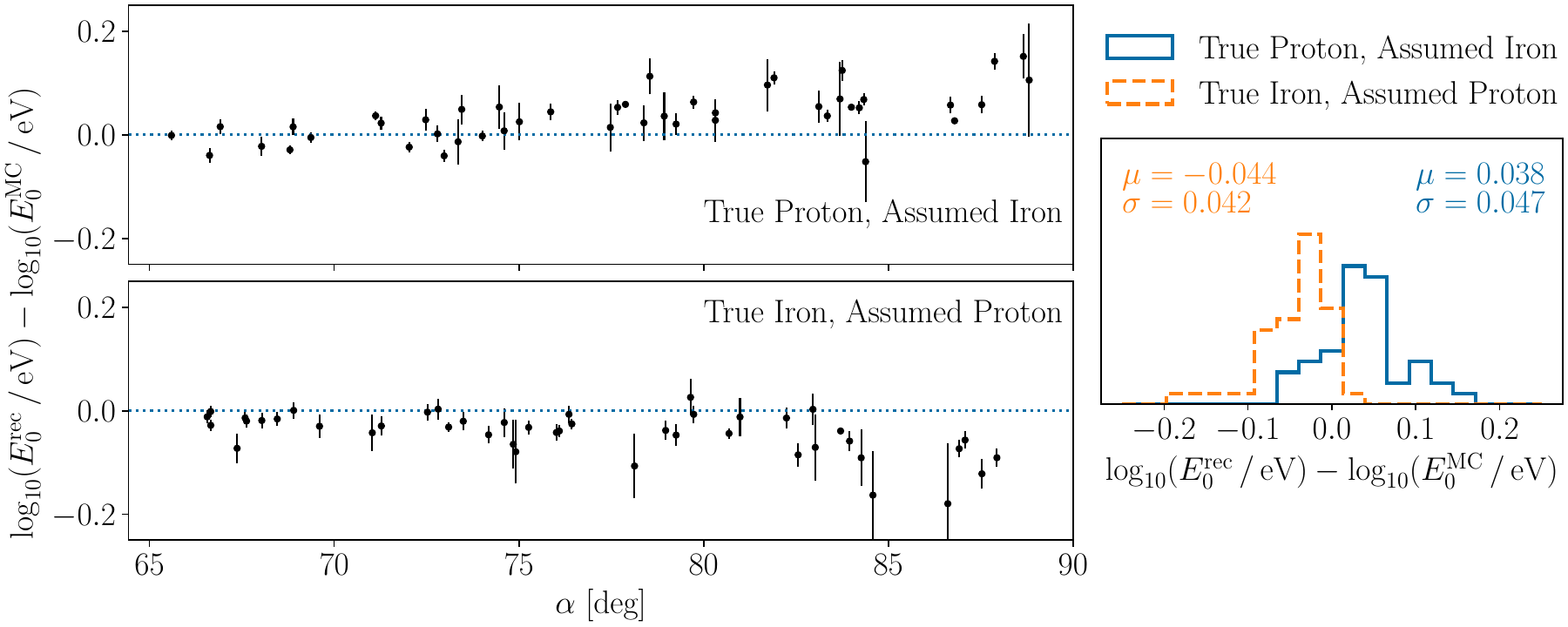}
    \caption{Results of the energy reconstruction method in the $100-450\unit{MHz}$ frequency range, applied to a subset of 50 simulated radio pulses produced by downward-going proton and iron air showers of several primary energies and incidence angles of radiation on the ice. Top left panel: Difference between the reconstructed $\log_{10}\left(E_0^{\rm rec}/\unit{eV}\right)$ and the true Monte Carlo energy $\log_{10}\left(E_0^{\rm MC}/\unit{eV}\right)$ as a function of the true incidence angle of radiation on the ice $\alpha$, for proton events reconstructed under the assumption of a primary iron. Bottom left panel: Same as the top panel, for iron events reconstructed under the assumption of a primary proton. Right panel: Distribution of biases in the reconstructed $\log_{10}\left(E_0/\unit{eV}\right)$ for all events, in the two scenarios of primary mass misidentification.}
    \label{fig:reco_reflex_massbias}
\end{figure*}

The impact of primary mass assumption in the reconstructed energies was studied following the same procedure employed for the case of direct events (Sec. \ref{sec:direct_massbias}). The energy of 50 proton- and 50 iron-induced air shower events with different geometries was reconstructed building the calibration libraries under the assumption of the opposite primary particle. The results of this estimation, performed without adding noise to the signals and using the true incidence angle on the surface $\alpha$ of each radio signal, is presented in Fig.\,\ref{fig:reco_reflex_massbias}.

As expected, a bias appears due to the unknown primary particle, $\mu_{\log\left(E\right)}\sim \pm0.04$ corresponding to a $\pm 10\%$ shift in the reconstructed primary energies in this worst-case scenario. Given the larger densities where downward-going air showers develop, the magnitude of the bias is smaller than in the case of direct events. In this case, the difference in the depth of $X_{\rm max}$ between iron and proton showers corresponds to a much smaller difference in the distance of shower maxima with respect to the detector.

Another significant effect is the growth of the bias in the reconstructed energies, with the incidence angle of the radiation. For highly inclined showers, the lower atmospheric density increases the distance between proton and iron shower maxima, making the primary-mass effect more pronounced than in more vertical showers. More detailed approaches, accounting for the evolution of cosmic ray composition with energy, could be used to reduce this uncertainty. 

\subsubsection{Frequency range and environmental noise}\label{sec:reflex_noisebias} 

\begin{figure}
    \centering
    \includegraphics[width=\linewidth]{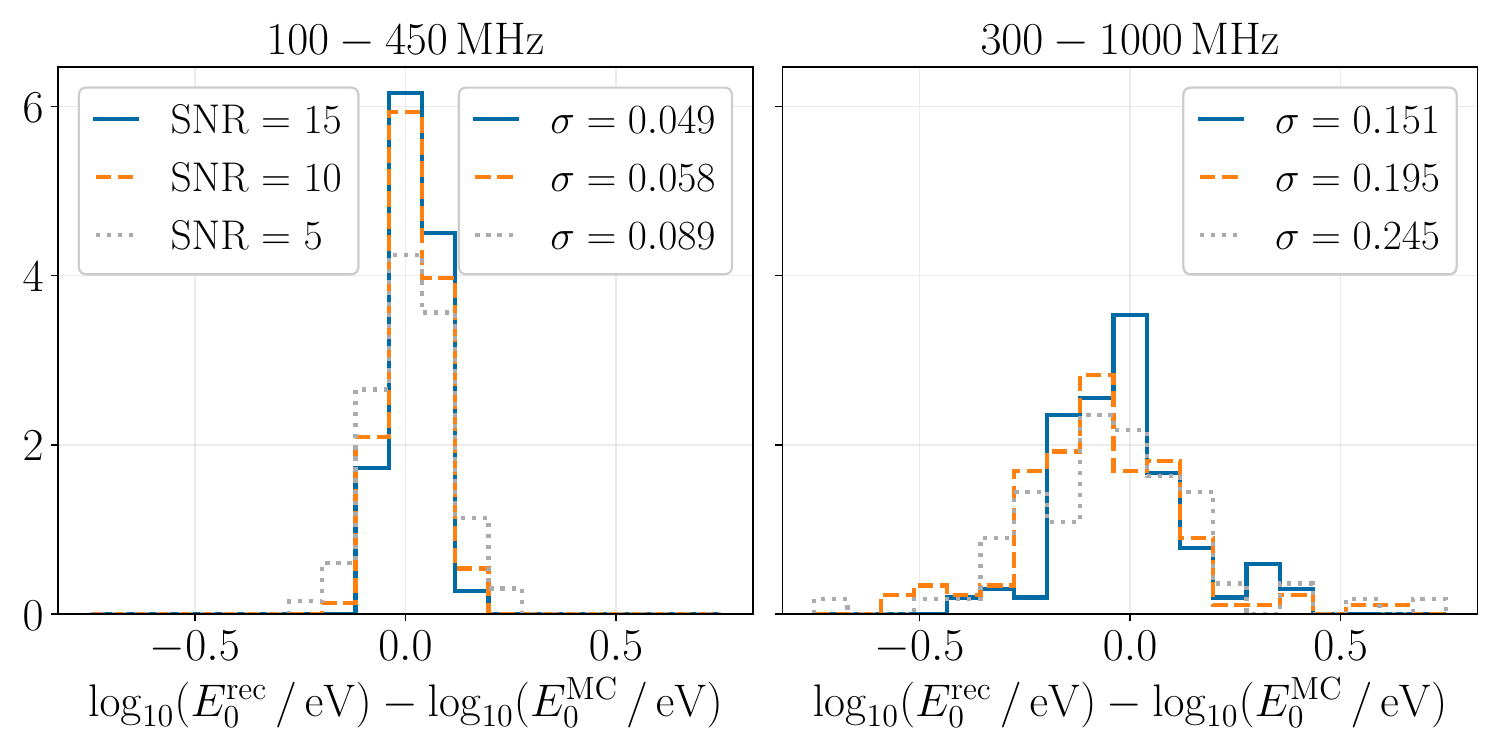}
    \caption{Left panel: Normalized distribution of the differences between the reconstructed energy $\log_{10}\left(E_0^{\rm rec}/\unit{eV}\right)$ and the true Monte Carlo energy $\log_{10}\left(E_0^{\rm MC}/\unit{eV}\right)$ for all events, after performing the spectral fits to Eq. \eqref{eq:expmodel} in the $100-450\unit{MHz}$ frequency range, under various assumptions for the SNR level. Right panel: Same as left panel, performing the spectral fits in the $300-1000\unit{MHz}$ frequency band. The standard deviation around the mean $\sigma\left(\log_{10}E\right)$ is indicated in both panels.}
    \label{fig:reco_reflex_noisebias}
\end{figure}

The effects of the SNR level in a more realistic scenario, as well as the frequency range where the measured signals are analyzed, was studied following the same procedure as in Sec. \ref{sec:direct_noisebias}. The treatment of the radio pulses after including gaussian noise, and the frequency ranges chosen to perform the energy estimation, are identical to those employed in the case of direct events. The results of the energy reconstruction in the presence of noise, under perfect mass knowledge and pointing resolution, are shown in Fig.\,\ref{fig:reco_reflex_noisebias}.

Similarly to the case of direct events, the energy resolution worsens when higher frequency components of the signal are included in the analysis, making the reconstruction more sensitive to event-to-event fluctuations on the position of shower maximum and to the background noise. These lack of biases indicate the possibility of performing an accurate energy reconstruction in the presence of background noise also for reflected events.

\section{Conclusions}\label{sec:conclusions}

Cosmic-ray-induced air showers are a guaranteed source of impulsive radio signals for experiments searching for UHE neutrinos. A thorough understanding of these signals is of great value for assessing detector performance and validating event reconstruction algorithms. For balloon-borne radio detectors in particular, specialized event reconstruction algorithms must be developed, as they must extract the shower properties from a single radio pulse received at a unique location.

In this work, we have reviewed and updated a simulation-based method for estimating the cosmic ray energy using radio signals recorded aboard balloons \cite{ANITAEnergyFlux}. This method exploits the dependence of the signal's frequency spectrum on the observer angular position to overcome the ambiguities in the shower axis direction. For the first time, we have adapted this method to reconstruct "direct" radio signals, produced by atmosphere-skimming cosmic ray air showers. While the logic of the method remains unchanged, the specific characteristics of the radio emission produced by these events \cite{RASPASS_Showers, RASPASS_Radio} required a new procedure to identify shower geometries compatible with the incoming direction of the signals. We demonstrate that the primary energy of these events can be reconstructed with a resolution comparable to that achieved for downward-going events (Fig.\,\ref{fig:reco_direct_nobias}). However, the reconstruction of direct events is very sensitive to the  instrument pointing resolution, as small changes in the shower geometry can change substantially the properties of the radio emission (Fig.\,\ref{fig:reco_direct_angularbias}). 
Furthermore, the energy resolution is sensitive to both the frequency range of the instrument and event-to-event fluctuations in $X_{\rm max}$ (Fig.\,\ref{fig:reco_direct_noisebias}). Because $X_{\rm max}$ strongly dictates the emission pattern, the unknown primary particle mass can induce a systematic bias as large as $\sim 25\%$ in the reconstructed energy (Fig.\,\ref{fig:reco_direct_massbias}). In practice,  this worst-case bias can be mitigated by utilizing a realistic, energy-dependent mixed-composition during the calibration of the reconstruction method with simulations.

We have also significantly updated the energy estimation method for reflected events by incorporating a detailed, track-by-track simulation of the radio reflection off a smooth spherical surface. Validation against simulated events demonstrates that this updated method provides an unbiased estimator of the primary energy, with an optimal resolution of the order of $\sim 11\%$, arising both from the ambiguity in the shower axis direction and event-to-event fluctuations (Fig.\,\ref{fig:reco_reflex_nobias}).

Unlike atmosphere-skimming showers, reflected events are remarkably robust against the pointing resolution uncertainties typical of current balloon payloads (Fig.\,
\ref{fig:reco_reflex_angularbias}).
Nevertheless, the unknown primary particle mass can also systematically bias the reconstructed energies, especially for the most inclined showers (Fig.\,\ref{fig:reco_reflex_massbias}). As with direct events, the ultimate energy resolution remains dependent on the frequency band of the instrument and the natural variations of individual showers around the expected average (Fig.\,\ref{fig:reco_reflex_noisebias}).

PUEO and POEMMA-Balloon with Radio are expected to record a sizable amount of cosmic-ray induced signals. This work establishes that robust energy reconstruction for both direct and reflected events is fully achievable for these missions. To maximize the accuracy of these reconstructions on actual flight data, future analyses must carefully account for instrumental pointing uncertainties and employ appropriate mass composition priors, particularly when evaluating highly inclined shower geometries.

\begin{acknowledgments}

This work has received financial support from:
Ministerio de Ciencia, Innovaci\'on y Universidades/Agencia Estatal de Investigaci\'on, MICIU/AEI /10.13039/501100011033, Spain
(PID2022-140510NB-I00, PCI2023-145952-2, RYC2019-027017-I, CNS2024-154676, and Mar\'\i a de Maeztu grant CEX2023-001318-M); Xunta de Galicia, Spain (CIGUS Network of Research Centers, and Consolidaci\'on ED431C-2025/11); and the European Union (Galicia Feder 2021-2027 Program). 

Part of the simulations presented in this paper were performed at the Clementina XXI supercomputer, part of the Sistema Nacional de Computación de Alto Desempeño de la República Argentina, Subsecretaría de Ciencia y Tecnología.

\end{acknowledgments}

\appendix
\renewcommand{\thesection}{\Alph{section}}

\counterwithin{figure}{section}
\counterwithin{equation}{section}

\renewcommand{\thefigure}{\thesection\arabic{figure}}
\renewcommand{\theequation}{\thesection.\arabic{equation}}

\section{Reflections on a spherical surface with \textsc{ZHAireS}}\label{app:reflex_mod}
\begin{figure}
    \centering
    \includegraphics[width=\linewidth]{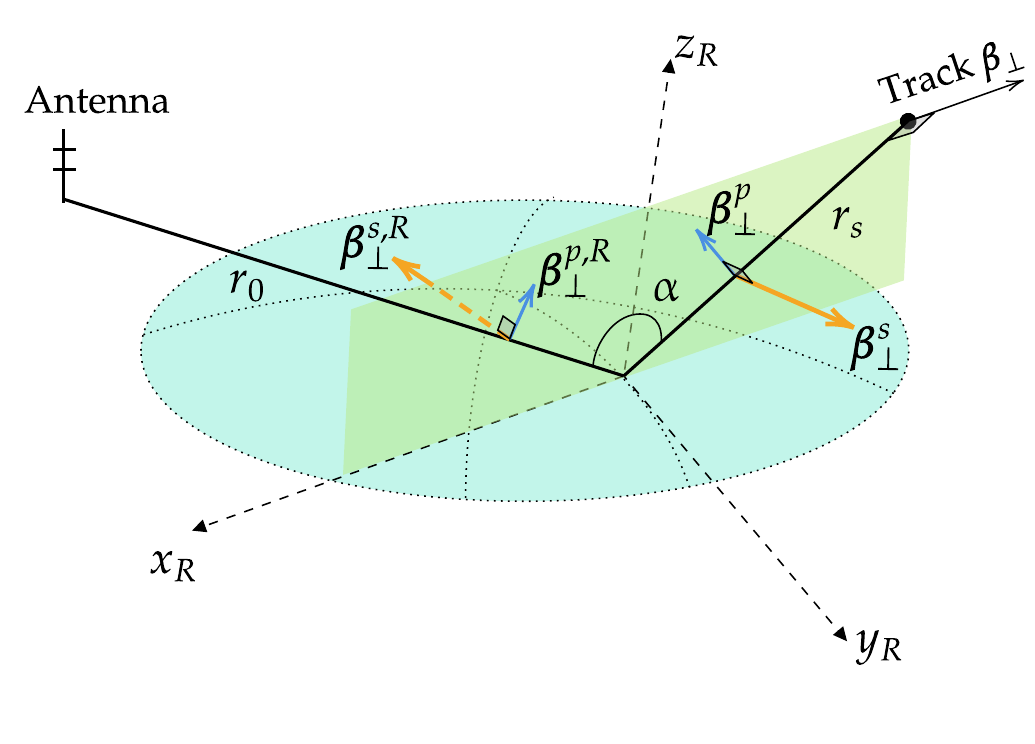}
    \caption{Treatment of reflection on a spherical surface in \textsc{ZHAireS}. The \textit{reflection plane} is indicated by the light green color, with the incidence angle $\alpha$ with respect to the local vertical direction $z_R$. The contribution of a particle track to the total electric field is proportional to the perpendicular velocity $\vect{\beta}_\perp$. The component of the electric field perpendicular to the reflection plane (proportional to $\vect{\beta}_\perp^s$, orange arrow) suffers a phase inversion at the reflection because the corresponding Fresnel coefficient $R_s<0$. In this sketch, the component of the electric field contained in the reflection plane (proportional to $\vect{\beta}_\perp^p$, blue arrow) does not suffer a phase inversion, corresponding to the case where the Fresnel coefficient $R_p>0$.}
    \label{fig:appreflexmod}
\end{figure}

The reflection of the emission from a single particle track in \textsc{ZHAireS} is sketched in Fig.\,\ref{fig:appreflexmod}, considering a spherical reflector of radius $R$. For two given points (in this case an antenna and a particle track), the reflection point on a spherical surface can be found by obtaining the roots of a quartic polynomial \cite{Eberly2008SphereReflections}. In the current implementation in \textsc{ZHAireS}, the polynomial root representing the spherical reflection point is found by means of a Newton-Raphson iterative method. 

In the ZHS formalism, the electric field emitted by a single particle track is proportional to the component of the particle velocity perpendicular to the line of sight to the antenna, $\vect{\beta}_\perp$ \cite{Zas:1991jv}. After the reflection point and the incidence angle $\alpha$ are found, this perpendicular velocity is expressed in a basis ($x_R$, $y_R$, $z_R$) containing the reflection plane (in a light green color in Fig.\,\ref{fig:appreflexmod}). Before calculating the contribution of the track to the total electric field, the component of $\vect{\beta}_\perp$ contained in the reflection plane ($\vect{\beta}_\perp^p$ in Fig.\,\ref{fig:appreflexmod}), and the component perpendicular to the reflection plane ($\vect{\beta}_\perp^s$) are corrected by the corresponding Fresnel coefficients $R_p$, $R_s$. The expressions for the Fresnel coefficients for reflection at the interface between two media of refractive index $n_1$, $n_2$ are given in Eq.\,\eqref{eq:Fresnel},
\begin{equation}
\begin{aligned}
    R_p &= \frac{n_2\cos\alpha-n_1\sqrt{1-\left(\frac{n_1}{n_2}\sin\alpha\right)^2}}{n_2\cos\alpha+n_1\sqrt{1-\left(\frac{n_1}{n_2}\sin\alpha\right)^2}}\;\;;\\\\
    R_s &= \frac{n_1\cos\alpha-n_2\sqrt{1-\left(\frac{n_1}{n_2}\sin\alpha\right)^2}}{n_1\cos\alpha+n_2\sqrt{1-\left(\frac{n_1}{n_2}\sin\alpha\right)^2}}\;.
\end{aligned}
\label{eq:Fresnel}
\end{equation}

Fresnel coefficients are only valid for reflections on a flat surface. A correction to the electric field produced by the particle track is now included in \textsc{ZHAireS}, in order to account for the decrease of signal amplitude due to divergence of rays after a reflection on a spherical surface. In the current implementation, this correction is derived from the Kirchhoff scalar diffraction formalism \cite{romerowolf2013, Gorham:2017xbo}. For an emitter of spherical waves of amplitude $\mathcal{A}$ placed at a position $\vect{r}_s$, and an observer at a position $\vect{r}_0$ (see Fig.\,\ref{fig:Kirchhoffintegral}), the received electric field for some polarization will have the shape: 

\begin{figure}
    \centering
    \includegraphics[width=\linewidth]{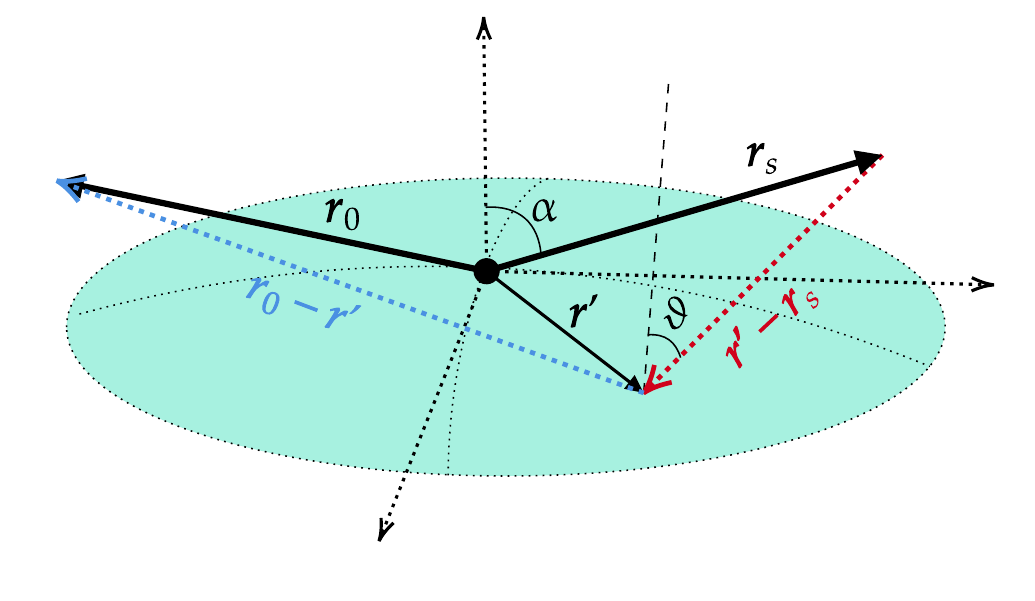}
    \caption{Notation in Eq. \eqref{eq:Kirchhoff}}
    \label{fig:Kirchhoffintegral}
\end{figure}

\begin{widetext}
\begin{equation}
    \mathcal{E}\left(\vect{r}_0\right)=\frac{k}{2\pi i}\int_{\rm sphere}d^2r'\,\mathcal{A}\,\frac{\exp\left(ikR'\right)}{R'}\,\mathcal{F}\left(\vect{r}',\vect{r}_0,\,\vect{r}_s\right)\,\frac{\exp\left(ikR\right)}{R}\,\cos\vartheta\left(\vect{r}',\vect{r}_0,\,\vect{r}_s\right)\,,\label{eq:Kirchhoff}
\end{equation}
\end{widetext}
where $R' = \left|\vect{r}'-\vect{r}_s\right|$ and $R = \left|\vect{r}_0-\vect{r}'\right|$, $\mathcal{F}$ represents the corresponding Fresnel coefficient, and $\vartheta$ is the local incidence angle. 

The typical distance between emitter/observer and the reflection point, will be of order $\mathcal{O}(10\unit{km})$ for the case of air showers, much larger than the wavelengths at radio frequencies relevant for this work. In the limit of small wavelength (i.e. $kR,\,kR'\gg 1$), this integral can be evaluated using the stationary phase approximation \cite{wong2001asymptotic, Gorham:2017xbo} around the specular reflection point between emitter and observer (indicated with a black dot in Fig.\,\ref{fig:Kirchhoffintegral}). Under this approximation, the integral \eqref{eq:Kirchhoff} reduces to \cite{Gorham:2017xbo}:
\begin{widetext}
\begin{equation}
\mathcal{E}\left(\vect{r}_0\right)\approx\underbrace{\frac{\mathcal{A}e^{ik(r_s+r_0)}}{r_s+r_0}\mathcal{F}(\alpha)}_{{\rm Flat}}\times\underbrace{\left[1 + \frac{2}{R}\left(\frac{1+\cos^2\alpha}{\cos\alpha}\right)\frac{r_0\,r_s}{r_0+r_s}+\frac{4}{R^2}\left(\frac{r_0\,r_s}{r_0+r_s}\right)^2\right]^{-1/2}}_{{\rm Correction}\;C_{\rm Defocus}}\;,
    \label{eq:app_defocus}
\end{equation}
\end{widetext}
where $R$ represents the radius of the reflecting spherical surface, and $\alpha$ is the specular reflection angle between emitter ($\vect{r}_s$) and observer ($\vect{r}_0$). In Eq.\,\eqref{eq:app_defocus} the first term corresponds to the electric field emitted by a single particle track in the ZHS algorithm, including the $1/R$ attenuation, the propagation phase and the Fresnel coefficient of reflection on a flat surface. Consequently, the second term represents the correction to the amplitude due to the reflection on a spherical surface, $C_{\rm Defocus}$. We evaluated numerically the integral \eqref{eq:Kirchhoff}, and checked its convergence to the analytical expression \eqref{eq:app_defocus} after the integration region is extended beyond the first Fresnel zone. The correction \eqref{eq:app_defocus} gives very similar results to some other expressions found in the literature, e.g. \cite{Prohira:2018mmv}.

\begin{figure}
    \centering
    \includegraphics[width=\linewidth]{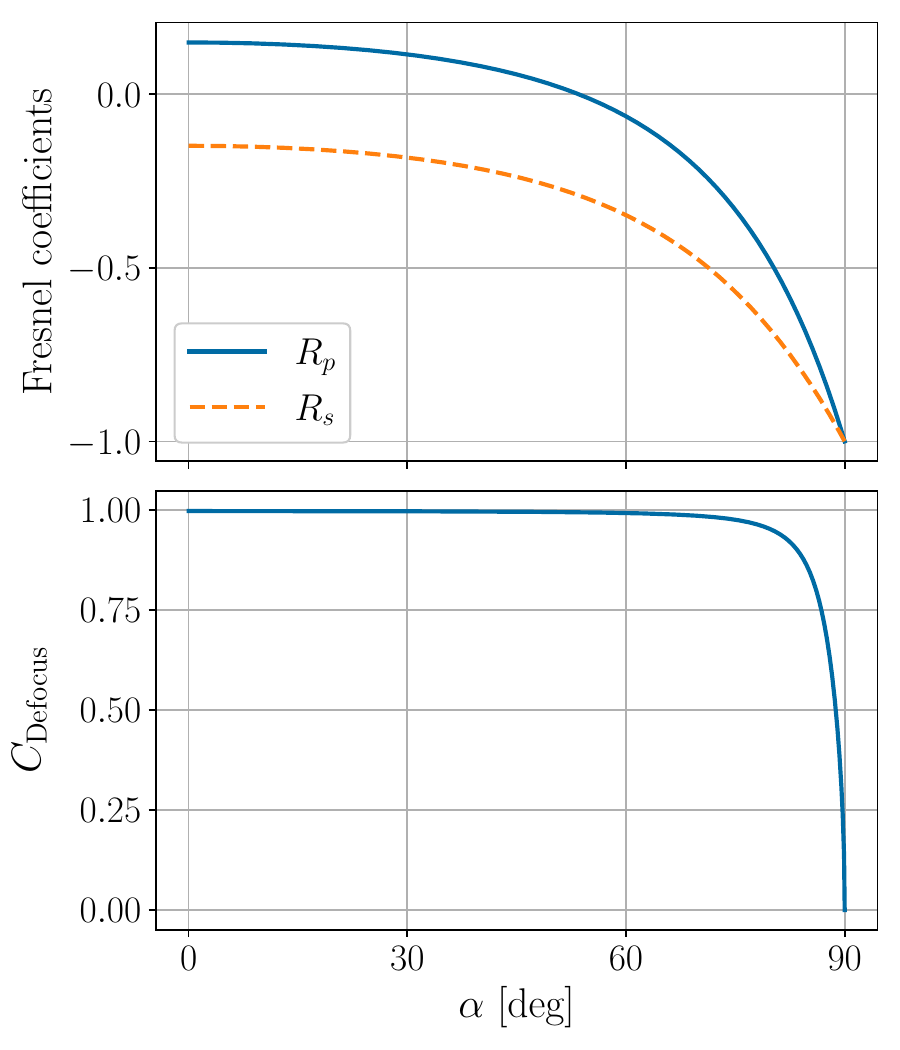}
    \caption{Top panel: Fresnel coefficients \eqref{eq:Fresnel} for reflection between two media of index of refraction $n_1=1.000325$ (air at sea level) and $n_2 = 1.35$ (ice). Bottom panel: Defocusing correction \eqref{eq:app_defocus} for the case of an emitter at an altitude of $15 \unit{km}$ and a receiver at an altitude of $36\unit{km}$ above the Earth's surface ($R = 6371\unit{km}$).}
    \label{fig:corrections}
\end{figure}

After applying Fresnel coefficients, and the correction \eqref{eq:app_defocus} to the emission of a particle track, its contribution to the total electric field received by an observer is computed, and the proper delay in time along the reflected ray path is applied. The dependence of Fresnel coefficients and $C_{\rm Defocus}$ on the incidence angle $\alpha$, is exemplified in Fig.\,\ref{fig:corrections}.

\begin{figure}
    \centering
    \includegraphics[width=\linewidth]{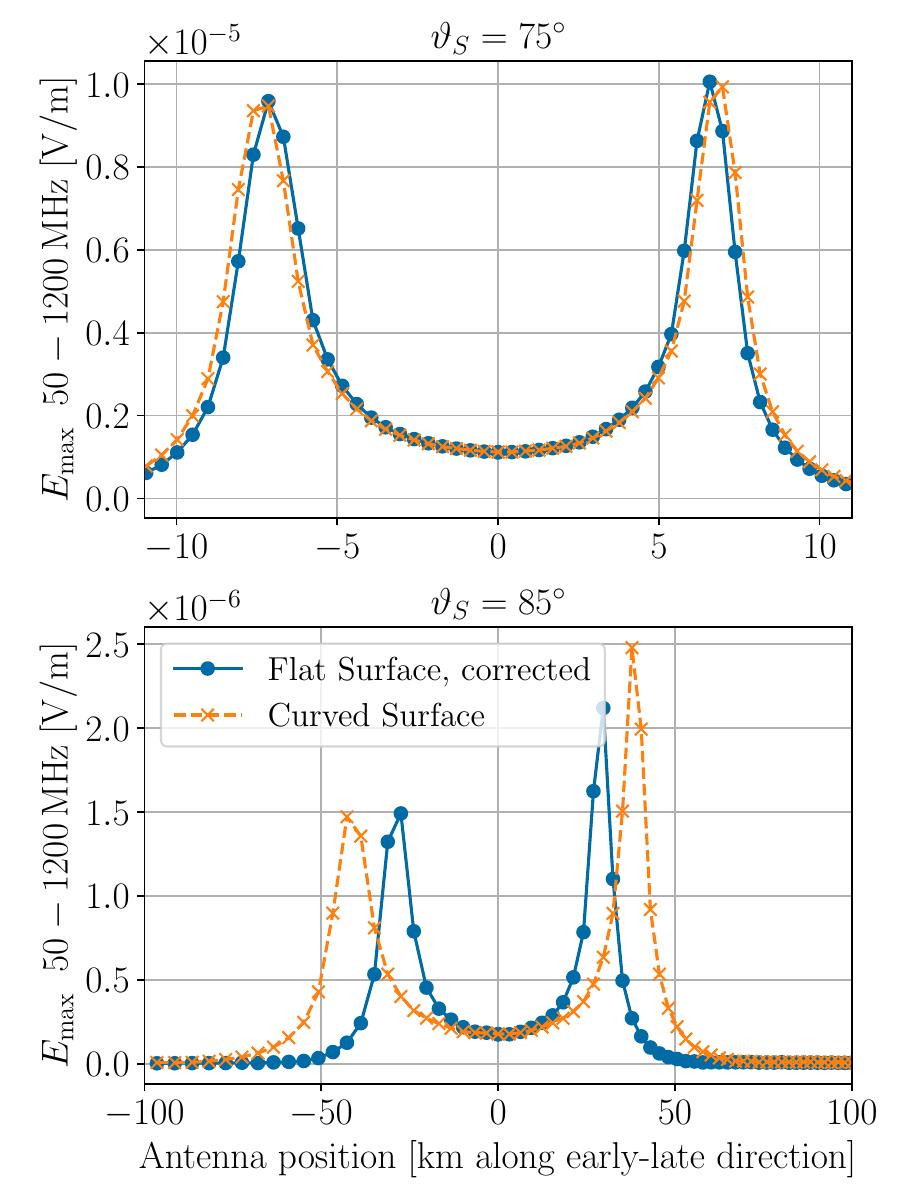}
    \caption{Lateral distribution of the peak electric field in the $50-1200\unit{MHz}$ band, observed at an altitude of $36\unit{km}$ along the early-late direction, after a reflection on either a flat (solid blue line) or spherical (dashed orange line) surface for a shower with zenith angle $\vartheta_S=75^\circ$ (left panel) and $\vartheta_S=85^\circ$ (right panel). Simulation results using a flat reflector are multiplied for the defocus term \eqref{eq:app_defocus} for comparison purposes.}
    \label{fig:zhaires_comparison}
\end{figure}

A comparison between \textsc{ZHAireS} versions, using either the previous flat surface approximation or the new implementation of reflections against a spherical surface, is shown in Fig.\,\ref{fig:zhaires_comparison}, where the emission of downward-going showers with different geometries is simulated twice, with the only difference being the shape of the reflector.

The geometric effects of a spherical reflecting surface are largely negligible for showers with zenith angles $\vartheta_S < 80^\circ$, as demonstrated in the left panel of Fig.\,\ref{fig:zhaires_comparison}. However, for more inclined showers, the Cherenkov cone illuminates a much larger area on the ground, causing the reflected radio footprint to broaden significantly. In this highly inclined regime, the signal amplitude at the detector altitude is also modified because the ray paths diverge differently off a curved surface compared to a flat one. Furthermore, at extreme inclinations ($\vartheta_S \gtrsim 87.5^\circ$), a fraction of the Cherenkov cone misses the Earth's spherical surface entirely, meaning the reflected signal no longer forms a \textit{closed} Cherenkov ring. By implementing these spherical reflection mechanics natively into the simulation framework, all of these complex effects are naturally accounted for in our energy reconstruction of highly inclined events.

\FloatBarrier
\bibliography{references}

@article{ANITAIV,
  title = {Unusual Near-Horizon Cosmic-Ray-like Events Observed by {ANITA-IV}},
  author = {Gorham, P. W. and others},
  collaboration = {ANITA},
  journal = {Phys. Rev. Lett.},
  volume = {126},
  issue = {7},
  pages = {071103},
  numpages = {7},
  year = {2021},
  month = {Feb},
  publisher = {American Physical Society},
  doi = {10.1103/PhysRevLett.126.071103},
  url = {https://link.aps.org/doi/10.1103/PhysRevLett.126.071103}
}

@article{HERON,
    author = "Kotera, Kumiko and others",
    collaboration = "GRAND, BEACON",
    title = "{The Hybrid Elevated Radio Observatory for Neutrinos (HERON) Project}",
    eprint = "2507.04382",
    archivePrefix = "arXiv",
    primaryClass = "astro-ph.IM",
    reportNumber = "PoS(ICRC2025)1078",
    doi = "10.22323/1.501.1078",
    journal = "PoS",
    volume = "ICRC2025",
    pages = "1078",
    year = "2025"
}

@article{ANITA:2008mzi,
    author = "Gorham, P. W. and others",
    collaboration = "ANITA",
    title = "{The Antarctic Impulsive Transient Antenna Ultra-high Energy Neutrino Detector Design, Performance, and Sensitivity for 2006-2007 Balloon Flight}",
    eprint = "0812.1920",
    archivePrefix = "arXiv",
    primaryClass = "astro-ph",
    doi = "10.1016/j.astropartphys.2009.05.003",
    journal = "Astropart. Phys.",
    volume = "32",
    pages = "10--41",
    year = "2009"
}

@article{POEMMA:2020ykm,
    author = "Olinto, A. V. and others",
    collaboration = "POEMMA",
    title = "{The POEMMA (Probe of Extreme Multi-Messenger Astrophysics) observatory}",
    eprint = "2012.07945",
    archivePrefix = "arXiv",
    primaryClass = "astro-ph.IM",
    doi = "10.1088/1475-7516/2021/06/007",
    journal = "JCAP",
    volume = "06",
    pages = "007",
    year = "2021"
}

@article{Schluter:2022mhq,
    author = {Schl{\"u}ter, Felix and Huege, Tim},
    title = "{Signal model and event reconstruction for the radio detection of inclined air showers}",
    eprint = "2203.04364",
    archivePrefix = "arXiv",
    primaryClass = "astro-ph.HE",
    doi = "10.1088/1475-7516/2023/01/008",
    journal = "JCAP",
    volume = "01",
    pages = "008",
    year = "2023"
}

@article{PierreAuger:2025jaw,
    author = "Abdul Halim, Adila and others",
    collaboration = "Pierre Auger",
    title = "{Results and plans to apply interferometry to air shower observations at the Pierre Auger Observatory}",
    eprint = "2509.04975",
    archivePrefix = "arXiv",
    primaryClass = "astro-ph.IM",
    reportNumber = "PoS ( ICRC2025 ) 386",
    doi = "10.22323/1.501.0386",
    journal = "PoS",
    volume = "ICRC2025",
    pages = "386",
    year = "2025"
}

@article{Guelfand:2025goo,
    author = "Guelfand, Marion and Decoene, Valentin and Martineau-Huynh, Olivier and Prunet, Simon and Tueros, Mat{\'\i}as and Macias, Oscar and Benoit-L{\'e}vy, Aur{\'e}lien",
    title = "{Reconstruction of inclined extensive air showers using radio signals: From arrival times and amplitudes to direction and energy}",
    eprint = "2504.18257",
    archivePrefix = "arXiv",
    primaryClass = "astro-ph.HE",
    doi = "10.1016/j.astropartphys.2025.103120",
    journal = "Astropart. Phys.",
    volume = "171",
    pages = "103120",
    year = "2025"
}

@article{Zhang:2025rzp,
    author = "Zhang, Kewen and Kaikai, Duan and Koirala, Ramesh and Tueros, Mat{\'\i}as and Zhang, Chao and Zhang, Yi",
    title = "{End-to-end reconstruction of ultra-high energy particle observables from radio detection of extensive air showers}",
    eprint = "2507.17266",
    archivePrefix = "arXiv",
    primaryClass = "astro-ph.IM",
    doi = "10.1140/epjc/s10052-025-15162-1",
    journal = "Eur. Phys. J. C",
    volume = "86",
    number = "1",
    pages = "11",
    year = "2026"
}

@article{PierreAuger:2016vya,
    author = "Aab, Alexander and others",
    collaboration = "Pierre Auger",
    title = "{Measurement of the Radiation Energy in the Radio Signal of Extensive Air Showers as a Universal Estimator of Cosmic-Ray Energy}",
    eprint = "1605.02564",
    archivePrefix = "arXiv",
    primaryClass = "astro-ph.HE",
    reportNumber = "FERMILAB-PUB-16-169-AD-AE-CD-TD",
    doi = "10.1103/PhysRevLett.116.241101",
    journal = "Phys. Rev. Lett.",
    volume = "116",
    number = "24",
    pages = "241101",
    year = "2016"
}

@article{Mitra:2020mza,
    author = "Mitra, P. and others",
    title = "{Reconstructing air shower parameters with LOFAR using event specific GDAS atmosphere}",
    eprint = "2006.02228",
    archivePrefix = "arXiv",
    primaryClass = "astro-ph.HE",
    doi = "10.1016/j.astropartphys.2020.102470",
    journal = "Astropart. Phys.",
    volume = "123",
    pages = "102470",
    year = "2020"
}

@article{ANITA:2008wdk,
    author = "Gorham, P. W. and others",
    collaboration = "ANITA",
    title = "{New Limits on the Ultra-high Energy Cosmic Neutrino Flux from the ANITA Experiment}",
    eprint = "0812.2715",
    archivePrefix = "arXiv",
    primaryClass = "astro-ph",
    reportNumber = "SLAC-PUB-14818",
    doi = "10.1103/PhysRevLett.103.051103",
    journal = "Phys. Rev. Lett.",
    volume = "103",
    pages = "051103",
    year = "2009"
}

@article{ANITA:2010hzc,
    author = "Gorham, P. W. and others",
    collaboration = "ANITA",
    title = "{Observational Constraints on the Ultra-high Energy Cosmic Neutrino Flux from the Second Flight of the ANITA Experiment}",
    eprint = "1003.2961",
    archivePrefix = "arXiv",
    primaryClass = "astro-ph.HE",
    doi = "10.1103/PhysRevD.82.022004",
    journal = "Phys. Rev. D",
    volume = "82",
    pages = "022004",
    year = "2010",
    note = "[Erratum: Phys.Rev.D 85, 049901 (2012)]"
}

@article{ANITA:2019wyx,
    author = "Gorham, P. W. and others",
    collaboration = "ANITA",
    title = "{Constraints on the ultrahigh-energy cosmic neutrino flux from the fourth flight of ANITA}",
    eprint = "1902.04005",
    archivePrefix = "arXiv",
    primaryClass = "astro-ph.HE",
    doi = "10.1103/PhysRevD.99.122001",
    journal = "Phys. Rev. D",
    volume = "99",
    number = "12",
    pages = "122001",
    year = "2019"
}

@article{ANITA:2018vwl,
    author = "Gorham, P. W. and others",
    collaboration = "ANITA",
    title = "{Constraints on the diffuse high-energy neutrino flux from the third flight of ANITA}",
    eprint = "1803.02719",
    archivePrefix = "arXiv",
    primaryClass = "astro-ph.HE",
    doi = "10.1103/PhysRevD.98.022001",
    journal = "Phys. Rev. D",
    volume = "98",
    number = "2",
    pages = "022001",
    year = "2018"
}

@article{Allison:2011wk,
    author = "Allison, P. and others",
    title = "{Design and Initial Performance of the Askaryan Radio Array Prototype EeV Neutrino Detector at the South Pole}",
    eprint = "1105.2854",
    archivePrefix = "arXiv",
    primaryClass = "astro-ph.IM",
    doi = "10.1016/j.astropartphys.2011.11.010",
    journal = "Astropart. Phys.",
    volume = "35",
    pages = "457--477",
    year = "2012"
}

@article{Abbasi:2025PO,
  author = "Abbasi, Rasha  and  others",
  title = "{Probing ultra-high-energy neutrinos with the IceCube-Gen2 in-ice radio array}",
  doi = "10.22323/1.501.1045",
  journal = "PoS",
  year = 2025,
  volume = "ICRC2025",
  pages = "1045"
}

@article{BEACON:2025qcq,
    author = "Zeolla, Andrew and others",
    collaboration = "BEACON",
    title = "{Sensitivity of BEACON to ultra-high energy diffuse and transient neutrinos}",
    eprint = "2504.13271",
    archivePrefix = "arXiv",
    primaryClass = "astro-ph.HE",
    doi = "10.1088/1475-7516/2025/09/033",
    journal = "JCAP",
    volume = "09",
    pages = "033",
    year = "2025"
}

@article{RNO-G:2020rmc,
    author = "Aguilar, J. A. and others",
    collaboration = "RNO-G",
    title = "{Design and Sensitivity of the Radio Neutrino Observatory in Greenland (RNO-G)}",
    eprint = "2010.12279",
    archivePrefix = "arXiv",
    primaryClass = "astro-ph.IM",
    doi = "10.1088/1748-0221/16/03/P03025",
    journal = "JINST",
    volume = "16",
    number = "03",
    pages = "P03025",
    year = "2021",
    note = "[Erratum: JINST 18, E03001 (2023)]"
}

@article{Horandel:2025Km,
  author = "Hörandel, Jörg",
  collaboration = "Pierre Auger",
  title = "{First Data of the 3000km2 Radio Detector at the Pierre Auger Observatory}",
  doi = "10.22323/1.501.0294",
  journal = "PoS",
  year = 2025,
  volume = "ICRC2025",
  pages = "294"
}

@article{ANITAIII,
  title = {Observation of an Unusual Upward-Going Cosmic-Ray-like Event in the Third Flight of {ANITA}},
  collaboration = "ANITA",
  author = {Gorham, P. W. and others},
  journal = {Phys. Rev. Lett.},
  volume = {121},
  issue = {16},
  pages = {161102},
  numpages = {6},
  year = {2018},
  month = {Oct},
  publisher = {American Physical Society},
  doi = {10.1103/PhysRevLett.121.161102},
  url = {https://link.aps.org/doi/10.1103/PhysRevLett.121.161102}
}

@article{Romero-Wolf:2014pua,
    author = "Romero-Wolf, A. and others",
    title = "{An interferometric analysis method for radio impulses from ultra-high energy particle showers}",
    doi = "10.1016/j.astropartphys.2014.06.006",
    journal = "Astropart. Phys.",
    volume = "60",
    pages = "72--85",
    year = "2015"
}

@article{Schoorlemmer:2020low,
    author = "Schoorlemmer, Harm and Carvalho, Washington R.",
    title = "{Radio interferometry applied to the observation of cosmic-ray induced extensive air showers}",
    eprint = "2006.10348",
    archivePrefix = "arXiv",
    primaryClass = "astro-ph.HE",
    doi = "10.1140/epjc/s10052-021-09925-9",
    journal = "Eur. Phys. J. C",
    volume = "81",
    number = "12",
    pages = "1120",
    year = "2021"
}

@article{ANITAEnergyFlux,
    author = "Schoorlemmer, H. and others",
    title = "{Energy and Flux Measurements of Ultra-High Energy Cosmic Rays Observed During the First ANITA Flight}",
    eprint = "1506.05396",
    archivePrefix = "arXiv",
    primaryClass = "astro-ph.HE",
    doi = "10.1016/j.astropartphys.2016.01.001",
    journal = "Astropart. Phys.",
    volume = "77",
    pages = "32--43",
    year = "2016"
}

@article{PBRTeam:2026wxs,
    author = "Adams, J. and others",
    collaboration = "PBR Team",
    title = "{POEMMA-Balloon with Radio: A multi-messenger, multi-detector balloon payload}",
    eprint = "2601.19997",
    archivePrefix = "arXiv",
    primaryClass = "astro-ph.IM",
    doi = "10.1088/1475-7516/2026/07/001",
    journal = "JCAP",
    volume = "07",
    pages = "001",
    year = "2026"
}

@article{AdamsJr:2026qlh,
    author = "Adams Jr., J. H. and others",
    title = "{The Extreme Universe Observatory on a Super-Pressure Balloon II: Mission, payload, and flight}",
    doi = "10.1016/j.astropartphys.2026.103263",
    journal = "Astropart. Phys.",
    volume = "182",
    pages = "103263",
    year = "2026"
}

@article{PUEOWhitePaper,
    author = {Qunicy Abarr and others},
    collaboration = "PUEO",
    title = "{The Payload for Ultrahigh Energy Observations (PUEO): a white paper}",
    eprint = "2010.02892",
    archivePrefix = "arXiv",
    primaryClass = "astro-ph.IM",
    doi = "10.1088/1748-0221/16/08/P08035",
    journal = "JINST",
    volume = "16",
    number = "08",
    pages = "P08035",
    year = "2021"
}

@article{auger:xmaxphaseI,
  title = {{Depth of maximum of air-shower profiles above ${10}^{17.7}\text{ }\text{ }\mathrm{eV}$ measured with the fluorescence detector of the Pierre Auger Observatory}},
  author = {Abdul Halim, A. and others},
  collaboration = {Pierre Auger Collaboration},
  journal = {Phys. Rev. D},
  volume = {114},
  issue = {4},
  pages = {043016},
  numpages = {28},
  year = {2026},
  month = {08},
  publisher = {American Physical Society},
  doi = {10.1103/n616-15v5},
  url = {https://link.aps.org/doi/10.1103/n616-15v5}
}

@article{PierreAuger:2025eun,
    author = "Abdul Halim, Adila and others",
    collaboration = "Pierre Auger",
    title = "{Energy Spectrum of Ultrahigh-Energy Cosmic Rays across Declinations -90{\textdegree} to +44.8{\textdegree} as Measured at the Pierre Auger Observatory}",
    eprint = "2506.11688",
    archivePrefix = "arXiv",
    primaryClass = "astro-ph.HE",
    doi = "10.1103/p4l5-hxlf",
    journal = "Phys. Rev. Lett.",
    volume = "135",
    number = "24",
    pages = "241002",
    year = "2025"
}

@article{Reflex,
    author = "Alvarez-Mu\~niz, Jaime and others",
    title = "{Simulations of reflected radio signals from cosmic ray induced air showers}",
    eprint = "1502.02117",
    archivePrefix = "arXiv",
    primaryClass = "astro-ph.HE",
    doi = "10.1016/j.astropartphys.2014.12.005",
    journal = "Astropart. Phys.",
    volume = "66",
    pages = "31--38",
    year = "2015"
}

@article{Martinelli_RiceMethod,
title = {Quantifying energy fluence and its uncertainty for radio emission from particle cascades in the presence of noise},
journal = {Astropart. Phys.},
volume = {168},
pages = {103091},
year = {2025},
issn = {0927-6505},
doi = {https://doi.org/10.1016/j.astropartphys.2025.103091},
url = {https://www.sciencedirect.com/science/article/pii/S0927650525000143},
author = {Sara Martinelli and others}
}

@article{Sibyll23d,
  title = {Hadronic interaction model {SIBYLL} 2.3d and extensive air showers},
  author = {Riehn, Felix and others},
  journal = {Phys. Rev. D},
  volume = {102},
  issue = {6},
  pages = {063002},
  numpages = {28},
  year = {2020},
  month = {Sep},
  publisher = {American Physical Society},
  doi = {10.1103/PhysRevD.102.063002},
  url = {https://link.aps.org/doi/10.1103/PhysRevD.102.063002}
}

@article{RASPASS_Showers,
doi = {10.1088/1475-7516/2024/07/065},
url = {https://dx.doi.org/10.1088/1475-7516/2024/07/065},
year = {2024},
month = {jul},
publisher = {IOP Publishing},
volume = {2024},
number = {07},
pages = {065},
author = {Tueros, Matías and others},
title = {Characterization of atmosphere-skimming cosmic-ray showers in high-altitude experiments},
journal = {JCAP}
}

@article{RASPASS_Radio,
    author = "Tueros, Mat{\'\i}as and Cabana-Freire, Sergio and {\'A}lvarez-Mu{\~n}iz, Jaime",
    title = "{Radio emission from atmosphere-skimming cosmic ray showers in high-altitude balloon-borne experiments}",
    eprint = "2409.13141",
    archivePrefix = "arXiv",
    primaryClass = "astro-ph.IM",
    doi = "10.1088/1475-7516/2025/01/112",
    journal = "JCAP",
    volume = "01",
    pages = "112",
    year = "2025"
}

@article{igrf14,
  author       = "{International Association of Geomagnetism and Aeronomy}",
  title        = "{IGRF-14}",
  doi = "10.5281/zenodo.14012303",
  journal = "Zenodo",
  year = "2024"
}

@article{PUEO_LF_ARENA24,
  author = "Ku, Yuchieh  and others",
  collaboration = "PUEO",
  title = "{The Low Frequency Instrument for the Payload for Ultrahigh Energy Observations (PUEO)}",
  doi = "10.22323/1.470.0019",
  journal = "PoS",
  year = 2024,
  volume = "ARENA2024",
  pages = "019"
}

@article{POEMMA_BR,
title = "{POEMMA-Balloon with Radio}: A balloon-born multi-messenger multi-detector observatory",
journal = {Nuclear Inst. and Methods in Physics Research A},
volume = {1069},
pages = {169819},
year = {2024},
issn = {0168-9002},
doi = {https://doi.org/10.1016/j.nima.2024.169819},
url = {https://www.sciencedirect.com/science/article/pii/S0168900224007459},
author = {Matteo Battisti and others}
}

@article{Prohira:2018mmv,
    author = "Prohira, S. and others",
    title = "{Antarctic surface reflectivity calculations and measurements from the ANITA-4 and HiCal-2 experiments}",
    eprint = "1801.08909",
    archivePrefix = "arXiv",
    primaryClass = "astro-ph.IM",
    doi = "10.1103/PhysRevD.98.042004",
    journal = "Phys. Rev. D",
    volume = "98",
    number = "4",
    pages = "042004",
    year = "2018"
}

@article{RadioAirShowers,
  title = {Coherent radiation from extensive air showers in the ultrahigh frequency band},
  author = {Alvarez-Mu\~niz, Jaime and Carvalho, Washington R. and Romero-Wolf, Andr\'es and Tueros, Mat\'{\i}as and Zas, Enrique},
  journal = {Phys. Rev. D},
  volume = {86},
  issue = {12},
  pages = {123007},
  numpages = {9},
  year = {2012},
  month = {Dec},
  publisher = {American Physical Society},
  doi = {10.1103/PhysRevD.86.123007},
  url = {https://link.aps.org/doi/10.1103/PhysRevD.86.123007}
}

@article{GRAND:2018iaj,
    author = "{\'A}lvarez-Mu{\~n}iz, Jaime and others",
    collaboration = "GRAND",
    title = "{The Giant Radio Array for Neutrino Detection (GRAND): Science and Design}",
    eprint = "1810.09994",
    archivePrefix = "arXiv",
    primaryClass = "astro-ph.HE",
    doi = "10.1007/s11433-018-9385-7",
    journal = "Sci. China Phys. Mech. Astron.",
    volume = "63",
    number = "1",
    pages = "219501",
    year = "2020"
}

@article{PierreAuger:2025rdo,
    author = "Abdul Halim, Adila and others",
    collaboration = "Pierre Auger",
    title = "{Measurement and Interpretation of UHECR Mass Composition at the Pierre Auger Observatory}",
    eprint = "2507.10292",
    archivePrefix = "arXiv",
    primaryClass = "astro-ph.HE",
    reportNumber = "PoS (ICRC2025) 331, PoS-ICRC2025-331",
    doi = "10.22323/1.501.0331",
    journal = "PoS",
    volume = "ICRC2025",
    pages = "331",
    year = "2025"
}

@misc{romerowolf2013,
      title={Concept and Analysis of a Satellite for Space-based Radio Detection of Ultra-high Energy Cosmic Rays}, 
      author={Andrew Romero-Wolf and Peter Gorham and Kurt Liewer and Jeffrey Booth and Riley Duren},
      year={2013},
      eprint={1302.1263},
      archivePrefix={arXiv},
      primaryClass={astro-ph.IM},
      url={https://arxiv.org/abs/1302.1263}, 
}

@article{Gorham:2017xbo,
    author = "Gorham, P. W. and others",
    title = "{Antarctic Surface Reflectivity Measurements from the ANITA-3 and HiCal-1 Experiments}",
    eprint = "1703.00415",
    archivePrefix = "arXiv",
    primaryClass = "astro-ph.IM",
    doi = "10.1142/S2251171717400025",
    journal = "J. Astron. Inst.",
    volume = "06",
    number = "02",
    pages = "1740002",
    year = "2017"
}

@article{ANITA:2010ect,
    author = "Hoover, S. and others",
    collaboration = "ANITA",
    title = "{Observation of Ultra-high-energy Cosmic Rays with the ANITA Balloon-borne Radio Interferometer}",
    eprint = "1005.0035",
    archivePrefix = "arXiv",
    primaryClass = "astro-ph.HE",
    doi = "10.1103/PhysRevLett.105.151101",
    journal = "Phys. Rev. Lett.",
    volume = "105",
    pages = "151101",
    year = "2010"
}

@techreport{Eberly2008SphereReflections,
  author      = {Eberly, David},
  title       = {{Alhazen's Problem: Reflection Point on a Sphere}},
  institution = {Geometric Tools},
  year        = {2008},
  url         = {https://www.geometrictools.com/Documentation/SphereReflections.pdf},
  note        = {Created February 1, 2008; last modified May 26, 2022. Accessed: 2026-06-05},
}

@book{wong2001asymptotic,
  title={Asymptotic approximations of integrals},
  author={Wong, R.},
  year={2001},
  publisher={SIAM},
  doi = "10.1137/1.9780898719260"
}

@article{Zas:1991jv,
    author = "Zas, E. and Halzen, F. and Stanev, T.",
    title = "{Electromagnetic pulses from high-energy showers: Implications for neutrino detection}",
    reportNumber = "MAD-PH-652",
    doi = "10.1103/PhysRevD.45.362",
    journal = "Phys. Rev. D",
    volume = "45",
    pages = "362--376",
    year = "1992"
}

\end{document}